\documentclass[article, a4paper, twoside]{memoir}

\counterwithout{section}{chapter}

\setsecnumdepth{subsection}

\setlrmarginsandblock{3.5cm}{2.5cm}{*}
\setulmarginsandblock{2.5cm}{*}{1}
\checkandfixthelayout

\usepackage[american]{babel}
\usepackage{csquotes}

\usepackage{amsmath,amssymb,amsthm}
\usepackage{bm,bbm}
\usepackage{mathtools}

\usepackage[dvipsnames, table]{xcolor}
\usepackage{graphicx}

\graphicspath{{./}}

\usepackage{tikz,pgfplots}
\usetikzlibrary{pgfplots.groupplots, arrows, positioning, calc, backgrounds}
\tikzset{>=latex} 
\pgfplotsset{compat=1.18}

\usepackage{enumitem}

\newcommand{\expect}[1]{ \mathbb{E} \left[ #1 \right] }

\newcommand{\cexpect}[2]{ \mathbb{E} \left[ #1 \: \big \vert #2 \: \right] }

\newcommand{\cproba}[2]{ \mathbb{P} \left( #1 \: \big \vert \: #2 \right) }

\newcommand{\setdef}[1]{ \left\{ #1 \right\} }

\newcommand{\indic}[1]{ \mathbbm{1}_{ \left\{ #1 \right\} } }

\newcommand{\indep}{\:\rotatebox[origin=c]{90}{$\models$}\:}

\DeclareMathOperator{\logit}{logit}

\newcommand{\target}[1]{\mathcal{T}_{#1}}

\usepackage{titling}

\title{Causal interpretation of the sibling comparison and its relation to the cross-over design}

\author{
  Simon Bang Kristensen$^{1, 2}$\thanks{
    Correspondance should be sent to Simon Bang Kristensen,
    Department of Public Health, Bartholins All\'{e} 2, 8000 Aarhus C, DK-Denmark,
    \href{mailto:simonbk@ph.au.dk}{simonbk@ph.au.dk}.
  }
  \and
  Christian Bressen Pipper$^{2, 3}$
  \and
  Jacob von Bornemann Hjelmborg$^{2}$ \\[1em]
  $^1$ \small Section for Biostatistics, Department of Public Health, Aarhus University\\
  $^2$ \small Department of Epidemiology, Biostatistics and Biodemography, University of Southern Denmark\\
  $^3$ \small Methods \& Outreach, Data Science, Novo Nordisk A/S
}

\newcommand{\shortauthor}{Kristensen et al.}

\newcommand{\paperkeywords}{Sibling comparisons; Twin data; Estimands; Cross-over design; Average causal treatment effect}

\date{
}

\usepackage[round]{natbib}
\usepackage{hyperref}

\hypersetup{
  colorlinks = true, linkcolor = black,
  citecolor = black, filecolor = black, urlcolor = blue,
  pdftitle = {\thetitle},
  pdfauthor = {\shortauthor},
  pdfkeywords = {\paperkeywords}
}

\begin{document}

\maketitle

\begin{abstract}
\noindent
The intuitive motivation for employing a sibling comparison design is to adjust for confounding that is constant within families. Such confounding can be caused by variables that otherwise might prove difficult or impossible to measure, for example factors relating to genetics or childhood environment and upbringing. Recent methodological investigations have shown that despite its intuitive appeal, the conventionally employed matched analysis of the sibling comparison does not relate to a well-defined causal target, even in the case of constant confounding. A main challenge is that the matched comparison will target the subpopulation of exposure discordant pairs. In the presence of an effect of the cosibling's exposure on the sibling's outcome, there is a second challenge to the interpretation of the matched analysis, namely that the effect corresponds to an intervention that always exposes the cosibling to the opposite exposure from the sibling. We aim to characterise the sibling comparison in terms of the celebrated cross-over design. Various estimands of interest are discussed before using this characterisation to establish more natural conditions for targeting an appropriate causal parameter. We cast the above-mentioned challenges of the sibling comparison in terms of those facing the cross-over trialist: that in order to target an appropriate estimand one must be able to argue the absence of a certain type of carry-over effect as well as a certain type of additivity, namely the absence of trial-by-treatment interaction, thus establishing that the former study design emulates the latter \emph{warts and all}. We explore weighting as a means to counter the effects of such interactions and to target other estimands. The weights rely on estimates of the unobserved confounding structure. Through simulations and an example analysis, we illustrate its potential usefulness as a supplement to assess the validity of the assumptions pertaining to the matched analysis. Lastly, we briefly discuss an extension of the weighting procedure to remove the final selection bias based on data from a population-level reference sample.
\end{abstract}

\section{Introduction}
\label{sec:introduction}

Sibling comparison designs have a long history for example in epidemiology, the basic rationale for their application being that by comparing individuals from the same family one is able to estimate exposure effects while controlling for a multitude of other factors that may otherwise have proven difficult to obtain, e.g. genetic and environmental variables. A particularly salient subtype of the sibling comparison considers pairs of twins, whereby one may compare individuals who are genetically similar or, for the case of monozygotic twins, genetically nearby identical. More recently, the apparent and intuitive benefits of the sibling comparison have been scrutinised more carefully from a methodological standpoint, especially using results from causal inference. In short, these investigations have sobered the classic enthusiasm for the sibling design. For instance, \cite{frisell_sibling_2012} show that the matched estimator may indeed suffer from worse confounding than the unmatched estimate in the presence of confounders that are not shared by the family members, thus invalidating the notion that the matched analysis must be superior since ``at least'' the shared confounding will be removed. This implies a trade-off between magnitude of bias stemming from the matched and the unmatched design. Moreover, if we think of the matched design as adjusting for the family, the role of the between-family factors becomes especially central since such factors may be interposed in the exposure-outcome relation \citep{sjolander_confounders_2017}.

Commonly, the within-family effect is of interest and any between-family effect is removed through the use of conditional estimation methods and this matched pair analysis constitutes the classic analysis strategy for the sibling-comparison design. \cite{sjolander_causal_2012} formally show that the conditional estimator is in fact an estimate of the causal effect confined to a subpopulation characterised by exposure discordance (i.e. a subpopulation where each sibling pair consists of an exposed and an unexposed sibling). This fact constitutes a substantial challenge to the interpretation of the sibling comparison, as results for this subpopulation may not generalise to the entire population, which is usually of interest. Other analysis strategies ignore the family structure by treating each sibling individually or include the cosibling's exposure as a covariate, the latter approach yielding equivalent estimates of the within effect as the matched analysis \citep{carlinRegressionModelsTwin2005,seaman_review_2014}.

The broader sense generalisability of the sibling comparison has similarly been disputed \citep{morleyCanWeGeneralise2005, sjolanderGeneralizabilityEffectMeasure2022}. The concern may be summarised as whether the estimated exposure effect may be generalised to a larger population than the one under study. There are at least two particularities to this question. Since the matched estimator has the above-mentioned conditional interpretation, one question would be to what extent this estimate generalises to the exposure concordant population. A second question is whether the effect generalises to all siblings aside from the ones under study -- a question particularly relevant when studying twins. For both questions, the answer of course depends on the disease area, exposures and outcomes under study. For measures such as the cumulative incidence and more elaborate biometric measures e.g. heritability of cancers twin studies generalizes well in the first sense, their being unconditional on concordance status \citep{skyttheCancerIncidenceMortality2019,mucciFamilialRiskHeritability2016a}, but their generalisability to a larger population in the second sense may be less obvious. However, for the matched case-cotwin studies measures obtained from conditioning on discordance of outcome are disputed and the general practice is to supplement these by reporting results of also the unmatched design \citep[e.g.][]{clemmensenTattooInkExposure2025}.

Another important concept, to be defined more precisely below, is what we term ``interference'', where the exposure of the cosibling may influence the outcome of the sibling. \cite{sjolander_carryover_2016} shows that interference effects will generally lead to biased estimates of the exposures effect on the outcome. A recent paper, \cite{petersen2020causal}, explores in detail the underlying causal assumptions of the sibling comparison in a setup allowing for interference of sibling exposures. Their results show that while the within-family effect obtained from sibling comparisons studies does have a causal interpretation, the interpretation is conditional on exposure discordance among the siblings and may, in the presence of interference effects, represent a rather oblique intervention on the exposures in which the cosibling is always subjected to the opposite exposure from the sibling.

This recently established catalogue of shortcomings pertaining to the sibling comparison as reviewed along with other statistical aspect in the recent paper \cite{sjolanderSiblingComparisonStudies2022} begs the question: under which assumptions does the comparison estimate a relevant, causal target? To put these assumptions into a more well-known framework we emulate the twin design to a cross-over design whereby we are able provide direct links to assumptions made in the context of a cross over design. Cross over trials are, as opposed to an observational study, typically highly controlled and modestly sized experiments whose participants are usually recruited from a highly selected group. Their kinship to the sibling comparison is noted for example by \cite{kalow_hypothesis_1998}, who argues that the cross-over trial may substitute for twin comparisons in the attempt to eliminate genetic variation from drug responses. As we will elaborate on below, the two types of studies overlap not only in their possibilities, but also in their pitfalls. In particular, two crucial assumptions of the crossover design coincide with the assumptions that make the sibling comparison work: Non-differential carryover and no trial-by-treatment interaction.

The article is structured as follows. We first establish the notation for the paper and introduce the basic statistical and causal assumptions. Our setup intentionally mimics that of \cite{petersen2020causal} and this first section also serves as a review of some of the results in this paper. We then define in detail possible target parameters for the sibling comparison and give a formal definition of interference. In order to formalise the comparison between the crossover trial and the sibling comparison, we then ask in which way the latter emulates the former as a target trial \citep[e.g.][]{hernanCausalInferenceWhat2020} and derive the relation between the non-differential carryover assumption and interference. We then turn to the problem of generalisability of the sibling comparison. Here, we expand the setup to investigate the generalisability of the sibling comparison with a view to the cross-over design and explore weighting as a means to reverse selection bias. The weights depend on unobserved confounders, but we propose to estimate these. The performance of the weighting procedure is explored in a simulation study and illustrated by an example analysis and we argue that it may be valuable as a method to estimate alternative targets or perform sensitivity analyses for the matched comparison. We end the article with a discussion where we extend the weighting procedure to be used in conjunction with reference data.

\section{Notation and background}
\label{sec:setup-notation}

Below we introduce the basic notation of the paper. The section also serves as a brief review of the results in \cite{petersen2020causal}.

\begin{figure}[htb]
  \centering
  \begin{tikzpicture}
    \node[] (U) {$U$};
    \node[right = of U, above of = U] (X1) {$X_1$};
    \node[right = of U, below of = U] (X2) {$X_2$};
    \node[right = of X1] (Y1) {$Y_1$};
    \node[right = of X2] (Y2) {$Y_2$};
    \draw[->] (X1) -> (Y1);
    \draw[->] (X2) -> (Y2);
    \draw[->] (X1) -> (Y2);
    \draw[->] (X2) -> (Y1);
    \draw[->, dashed] (U) -- (X1);
    \draw[->, dashed] (U) -- (X2);
    \draw[dashed] (U.90) edge[bend left = 60, ->] (Y1.90);
    \draw[dashed] (U.270) edge[bend right = 60, ->] (Y2.270);
  \end{tikzpicture}
  \caption{DAG representing the assumed relationship between the
    outcomes $Y$, exposures $X$ and unobserved confounders $U$. The
    potentially unmeasured effect is denoted by the dashed arrow.}
  \label{fig:dag-assumed}
\end{figure}
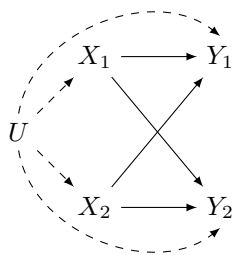

We denote by $\left(X_1, X_2, Y_1, Y_2 \right)$ the observations on a single family where $X_j \in \setdef{0, 1}$ is the observed, binary exposure status of the $j$'th sibling and $Y_j \in \mathbb{R}$ is the associated outcome. The data consists of $N$ such observations $\setdef{\left(X_i, Y_i\right)}_{i=1}^N$. Moreover, let $U$ be a vector of, possibly unobserved, potential confounders. We assume the interrelationship between the variables to be given by the directed acyclic graph \citep[DAG][]{greenland_causal_1999} in Figure \ref{fig:dag-assumed}. Causal effects are discussed is terms of counterfactual outcomes. Denote by $\left(Y_1(x_1, x_2), Y_2(x_1, x_2) \right)$ the outcome of a family if, possibly contrary to the fact, the exposure on sibling one were set to $x_1$ and the exposure on sibling two to $x_2$. When $W$ is some random variable, we write $f_W (\cdot)$ for its density function.

As in Figure \ref{fig:dag-assumed} we will in the following use DAGs to encode assumptions on conditional independences in observed and unobserved variables. We will also use design schematics, a less formal type of graphs, to represent treatment arms in trial designs. Schematics use rectangular nodes in boxes and a graph is identified as a DAG or schematic in the figure legend.

We refer to the situation in Figure \ref{fig:dag-assumed} where the exposure of a sibling may affect the outcome of the cosibling as \emph{interference}. The phenomenon has in the literature been referred to both as ``carry-over'' \citep{sjolander_carryover_2016} and ``cross-over'' \citep{petersen2020causal} effects, but since we will discuss cross-over designs below, we prefer to reserve this terminology and follow \cite{hernanCausalInferenceWhat2020} in their terminology of ``interference''. Our focus is on interference of a sibling's outcome from a cosiblings exposure. An alternative point of interest is the effect of a cosibling's exposure on the sibling's exposure, a situation which is implicitly contained in the DAG in Figure \ref{fig:dag-assumed} through a correlation between $X_1$ and $X_2$ caused by a possibly unmeasured common ancestor in $U$, also see the discussion below.

As in \cite{petersen2020causal} we impose the following statistical assumptions.
\begin{itemize}
\item[(S1)] The distribution of observations on siblings is symmetric within a family, i.e.
  \begin{equation}
    \label{eq:3}
    \left(X_1, X_2, Y_1, Y_2\right) \sim \left(X_2, X_1, Y_2, Y_1\right) .
  \end{equation}
\item[(S2)] The observations on two different families $i \not= i'$ are independent, i.e.
  \begin{equation}
    \label{eq:2}
    \left(X_{i1}, X_{i2}, Y_{i1}, Y_{i2} \right) \indep \left(X_{i'1}, X_{i'2}, Y_{i'1}, Y_{i'2} \right) .
  \end{equation}
\end{itemize}
Note that the second assumption in \cite{petersen2020causal} is that the families be i.i.d., which conflicts with the model that will be proposed shortly (in which the families have different means).

Consider the fixed effects model,
\begin{equation}
  \label{eq:1}
  M:\quad
  Y_{ij} = a_i + \beta^w X_{ij} + \epsilon_{ij} , \quad
  \text{for } j = 1,2, \text{ and } i = 1, \ldots, N .
\end{equation}
Here $\beta^w$ is the within-family effect of the exposure, which is typically taken to be the parameter of interest, while $a_i$ denotes a family-specific intercept. Further, the error terms $\epsilon_{ij}$ are assumed to be i.i.d. Noting that the deviation from the family mean is,
\begin{equation}
  \label{eq:4}
    Y_{ij} - \bar{Y}_{i.}=  \beta^w \left(X_{ij} - \bar{X}_{i.} \right) + \tilde{\epsilon}_{ij} , 
\end{equation}
for homoscedastic errors $\tilde{\epsilon}_{ij}$, leads to the usual OLS estimator of $\beta^w$. The OLS estimator in the model (\ref{eq:4}) is in fact the conditional likelihood estimator in the fixed effects model (\ref{eq:1}), where the family-specific intercepts have been removed. \citeauthor{petersen2020causal} show that,
\begin{equation}
  \label{eq:5}
  \hat{\beta}^w_{\text{OLS}} \overset{a.s.}{\longrightarrow}
  \cexpect{Y_1}{X_1 = 1, X_2 = 0} - \cexpect{Y_1}{X_1 = 0, X_2 = 1} ,
\end{equation}
as the number of families $N$ goes to infinity.

The discussion is now expanded to one of causality under the potential outcome framework. Recall that $Y_j(x_1, x_2)$ denotes the outcome on sibling $j$ when, possibly contrary to the fact, the first sibling's exposure is fixed at $x_1$ and the second sibling's exposure is fixed at $x_2$. The following causal assumptions are imposed.
\begin{itemize}
\item[(C1)] A set of, possibly unmeasured, variables $U$ are sufficient to control for confounding so that \mbox{$Y_j(x_1, x_2) \indep \left(X_1, X_2\right) \:\big\vert\: U$} for both $j=1,2$. Further, assume that the distribution of these variables in a population with a given exposure status coincides with the distribution when the exposure status of the two siblings is reversed, 
  \begin{equation}
    \label{eq:6}
    \cproba{U \in \cdot}{X_1 = x_1, X_2 = x_2} = \cproba{U \in \cdot}{X_1 = x_2, X_2 = x_1}
  \end{equation}
  for all possible values $x_1, x_2$.
\item[(C2)] The counterfactuals obey consistency in the sense that
  \begin{equation}
    \label{eq:7}
    \cproba{Y_j(x_1, x_2) = Y_j}{X_1 = x_1, X_2 = x_2} = 1 ,
  \end{equation}
  for $j=1,2$.
\end{itemize}

Under these causal assumptions, \citeauthor{petersen2020causal} show that,
\begin{equation}
  \label{eq:9}
  \begin{aligned}
    &\cexpect{Y_1}{X_1 = 1, X_2 = 0} - \cexpect{Y_1}{X_1 = 0, X_2 = 1} \\
    &= \cexpect{Y_1(1,0) - Y_1(0,1)}{X_1 = 1, X_2=0} \\
    &= \cexpect{Y_1(1,0) - Y_1(0,1)}{X_1 = 0, X_2=1} ,
  \end{aligned}
\end{equation}
where we recognise the left-hand side as the limit of the OLS estimator of the within family exposure effect. Adding and subtracting $\cexpect{Y_1(0,0)}{X_1 \not = X_2}$, we follow \citeauthor{petersen2020causal} in concluding that,
\begin{equation}
  \label{eq:10}
  \hat{\beta}^w_{\text{OLS}} \longrightarrow
  \cexpect{Y_1(1,0) - Y_1(0,0)}{X_1 \not= X_2} - \cexpect{Y_1(0,1) - Y_1(0,0)}{X_1 \not= X_2} .
\end{equation}
When there is no interference effect (as we will make more precise below) we would expect the expectation of the first sibling's outcome not to depend on the cosibling's exposure so that the second expectation on the right-hand side of the convergence is zero and thus the effect is comparing an exposed sibling one to an unexposed sibling one. However, the effect is still conditional to the population of exposure discordant families.

In summary, we note the following two shortcomings of the effect estimated by the matched analysis

\begin{enumerate}
\item It represents an unconventional intervention
\item It is conditional to the subpopulation of exposure-discordant families
\end{enumerate}

As \citeauthor{petersen2020causal} note, these shortcomings do not imply that the sibling comparison will be completely without merit in all situations. Indeed, they identify scenarios, where the effect may be attached a meaningful interpretation, for example when the interference effect can be argued to be so small as to be inconsequential. Below, we extend and formalise these arguments and relate the involved assumptions to those driving the cross-over design.

\section{Causal targets and interference}
\label{sec:targets-estimation}

When interpreting statistical effects, a common strategy is to start with a model and inquire about the interpretation of one or more of its parameters. In contrast to this model-first approach, a target-first approach would pose a target parameter for estimation, so that a model simply represents a tool for estimation, and several different models may be relevant for estimating the target under different assumptions. This latter approach has received increased attention, both through the advances of causal inference and from clinical trialists, where the target is known as an \emph{estimand} \citep[see the addendum to the statistical efficacy guidelines in technical report][]{emachmpich4362212017_ich_nodate}. To avoid connotations relating to the regulatory setting, we will prefer the term \emph{causal target} for the remainder of the paper.

The above-mentioned developments also reveal the fact that actually defining a target is often challenging in itself, and different targets may be of interest to different stakeholders, in the wide sense, of the study. Below we consider three targets that we would argue could be of interest in a sibling comparison, where interference is expected. In the absence of interference, the targets all reduce to the well known average treatment effect. We do not claim that the list of targets is exhaustive, and we return to some possible modifications of the target in the discussion. The targets are formulated focusing on the first sibling, but the formulation would proceed analogously for the second, and under certain assumptions of symmetry (such as those considered above) the distinction is irrelevant. Consider the following target parameters,
\begin{enumerate}
\item $\target{1} = \expect{Y_1(1, X_2) - Y_1(0, X_2)}$. A sibling-level intervention on the first sibling only, i.e. the total effect of exposure on sibling one.
\item $\target{2} (x) = \expect{Y_1(1, x) - Y_1(0, x)}$ for $x= 0,1$. A sibling-level intervention effect, that is similar to the notion of a direct controlled effect, where the cosibling's exposure status is fixed to exposure ($x=1$) or non-exposure ($x=0$).
\item $\target{3} = \expect{Y_1(1, 1) - Y_1(0, 0)}$. A family-level intervention effect assigning both siblings to either exposure or non-exposure.
\end{enumerate}

The causal targets introduced above may be estimated from a variety of different designs, but it may be illustrative to note how they correspond naturally to different target trials. The target parameter $\target{1}$ may be interpreted as follows. A two-armed parallel randomised trial is performed, randomising first siblings to the two arms $X=1$ and $X=0$ essentially ignoring the family-structure of the data. Afterwards, we analyse all first siblings from the families, and the difference in averages between the two arms is an estimate of $\target{1}$. Similarly, we may think of $\target{3}$ as arising from a two-arm cluster-randomised design, where families are randomised to exposure or non-exposure (i.e. the two arms $(X_1,X_2) = (1,1)$ and $(X_1,X_2) = (0,0)$). After the trial is completed, the differences in average outcome in first siblings from exposed families minus the average from unexposed families estimates $\target{3}$. The direct controlled effect may for example be thought of as effects in a stratified randomisation setup, where sibling one is randomised in strata defined by the exposure status of sibling two. The treatment contrast in the two strata $X_2 = 0$ and $X_2 = 1$ would then estimate $\target{2}(0)$ and $\target{2}(1)$, respectively.

We now give a more formal definition of the notion of interference discussed above. We consider the following two definitions of no interference.
\begin{enumerate}
\item If $\expect{Y_1(x_1, x_2)} = \expect{Y_1(x_1, 1 - x_2)}$ for any $x_1, x_2 \in \setdef{0, 1}$, we say that there is no interference in the strong sense.
\item If $\target{2}(1) - \target{2}(0) = 0$, we say that there is no interference in the weak sense.
\end{enumerate}
Often \citep[e.g. in][]{petersen2020causal}, interference is discussed in terms of the ``strong'' definition. We note that the strong definition implies the weak. We will further introduce the following concept,
\begin{enumerate}[resume]
\item If $\expect{Y_1 (1, 1) - Y_1 (1, 0)} = \expect{Y_1 (0, 0) - Y_1 (0, 1)} =: \tau$, we say that there is non-differential interference (NDI).
\end{enumerate}
NDI expresses that the interference obtained by removing the cosibling's exposure in an exposed family coincides with the interference arising from adding exposure to the cosibling in an unexposed family.
Note that under NDI,
\begin{equation}
  \label{eq:27}
  \begin{aligned}
    \target{2}(1) - \target{2}(0)
    &=
    \expect{Y_1(1,1) - Y_1(0,1) - \left(Y_1(1,0) - Y_1(0,0) \right)} \\
    &=
    2 \tau .
  \end{aligned}
\end{equation}
This in turn means that if NDI holds, then no interference in the weak sense will imply $\tau = 0$, and thus,
\begin{equation}
  \label{eq:28}
  \expect{Y_1 (1, 1)} = \expect{Y_1 (1, 0)}, \quad
  \text{and} \quad
  \expect{Y_1 (0, 0)} = \expect{Y_1 (0, 1)} ,
\end{equation}
or, in other words, that there is no interference in the strong sense. We could also say that in a world where NDI holds, no interference in the weak and strong sense are equivalent.

We also have that,
\begin{equation}
  \label{eq:29}
  \begin{aligned}
    \target{3}
    &=
      \expect{Y_1(1, 1) - Y_1(0, 1)} + \expect{Y_1(1, 0) - Y_1(0, 0)} + \expect{Y_1(0, 1) - Y_1(1, 0)} \\
    &=
      \target{2}(1) + \target{2}(0) - \target{3}  + \expect{Y_1(1, 1) - Y_1(0, 0) - Y_1(1,0) + Y_1(0,1)},
  \end{aligned}
\end{equation}
or,
\begin{equation}
  \label{eq:30}
  \target{3} = \frac{1}{2} \left( \target{2}(1) + \target{2}(0) \right) +
  \frac{1}{2} \left\{\expect{Y_1(1, 1) - Y_1(0, 0) - Y_1(1,0) + Y_1(0,1)} \right\} ,
\end{equation}
where the term in the curly brackets may be interpreted as the expected deviation from non-differential interference. Thus, under NDI,
\begin{equation}
  \label{eq:25}
  \target{3} \overset{NDI}{=} \frac{1}{2} \left( \target{2}(1) + \target{2}(0) \right) ,
\end{equation}
so that we may obtain the family-level effect as the simple average of the two controlled direct effects.

\section{A cross-over target trial}
\label{sec:target-trials}

The interpretation of a clinical randomised trial is often perceived as more straightforward than that of an observational study. The target trial concept \citep[e.g.][]{hernanCausalInferenceWhat2020} invites researchers to consider the observational study as an emulation of a clinical trial. By specifying this trial in sufficient detail it becomes apparent in which ways the observational data may imitate the trial and in which ways the two diverge. As an example of the latter the nature of the observational study will usually invalidate the randomisation of the target trial. We proceed in a general fashion below and as such will employ the target trial concept mainly as an instrument of exposition.

\cite{kalow_hypothesis_1998} noted the analogy between the twin-comparison and the cross-over design. They discuss using the cross-over to emulate a twin comparison in pharmacological experiments which forms the basis of a classic Fisherian genetic components of variance analysis. We will proceed with the converse and emulate the cross-over trial by the sibling comparison to expand on the causal interpretation of the latter and in particular illustrate how the concept of interference can be represented in a within-family cross-over design.

\begin{figure}[htb]
  \centering
  \begin{tikzpicture}
    \tikzstyle{nsty} = [
    draw,
    rectangle,
    fill = gray!10,
    minimum width = {width("$Y_{j2}(x,x)$") + 2pt},
    text width = {width("$Y_{j2}(x,x)$") - 2pt},
    minimum height = 1cm]
    \tikzstyle{nsty2} = [
    draw,
    rectangle,
    minimum width = {width("$Y_{j2}(x,x)$") + 2pt},
    text width = {width("$Y_{j2}(x,x)$") - 2pt},
    minimum height = 1cm]
    \node (selection) {Selection};
    \node[above right=2.2cm and 0.5cm of selection] (alph_i) {$\alpha_i$};
    \node[below right=2.2cm and 0.5cm of selection] (alph_j) {$\alpha_j$};
    \draw[->, dashed] (selection) -- (alph_i);
    \draw[->, dashed] (selection) -- (alph_j);
    \node[right = of alph_i] (rando) {Randomisation};
    \draw[->, dashed] (alph_i) -- (rando);
    \node[nsty, above right = of rando] (arm1t1) {$Y_{i1}(1, 0)$};
    \node[nsty, below right = of rando] (arm2t1) {$Y_{i2}(1, 0)$};
    \node[nsty2, right = of arm2t1] (arm1t2) {$Y_{i1}(0, 1)$ $ + \pi + \lambda_1$};
    \node[nsty2, right = of arm1t1] (arm2t2) {$Y_{i2}(0, 1)$ $ + \pi + \lambda_0$};
    \node[above = of arm1t1] (p1) {Period 1};
    \node[above = of arm2t2] (p2) {Period 2};
    \draw[->, dashed] (rando) -- (arm1t1);
    \draw[->, dashed] (rando) -- (arm2t1);
    \draw[->] (arm1t1) -- (arm1t2);
    \draw[->] (arm2t1) -- (arm2t2);
    \node[right = of alph_j] (rando) {Randomisation};
    \draw[->, dashed] (alph_j) -- (rando);
    \node[nsty, above right = of rando] (arm1t1j) {$Y_{j1}(0, 1)$};
    \node[nsty, below right = of rando] (arm2t1j) {$Y_{j2}(0, 1)$};
    \node[nsty2, right = of arm2t1j] (arm1t2j) {$Y_{j1}(1, 0)$ $ + \pi + \lambda_0$};
    \node[nsty2, right = of arm1t1j] (arm2t2j) {$Y_{j2}(1, 0)$ $ + \pi + \lambda_1$};
    \draw[->, dashed] (rando) -- (arm1t1j);
    \draw[->, dashed] (rando) -- (arm2t1j);
    \draw[->] (arm1t1j) -- (arm1t2j);
    \draw[->] (arm2t1j) -- (arm2t2j);
    \draw[thick,dotted] ($(arm2t2.north west)+(-0.3,0.6)$)  rectangle ($(arm1t2j.south east)+(0.3,-0.6)$);
  \end{tikzpicture}
  \caption{Schematic of a within-family ABBA cross-over design, as
    described in the text. The period effect is represented by $\pi$
    and $\lambda_0$ and $\lambda_1$ is the potential carry-over effect
    of the exposures from the first to the second period. The gray
    nodes are unobserved, while the unfilled nodes in the dotted box
    are observed.}
  \label{fig:schematic-cross-over}
\end{figure}
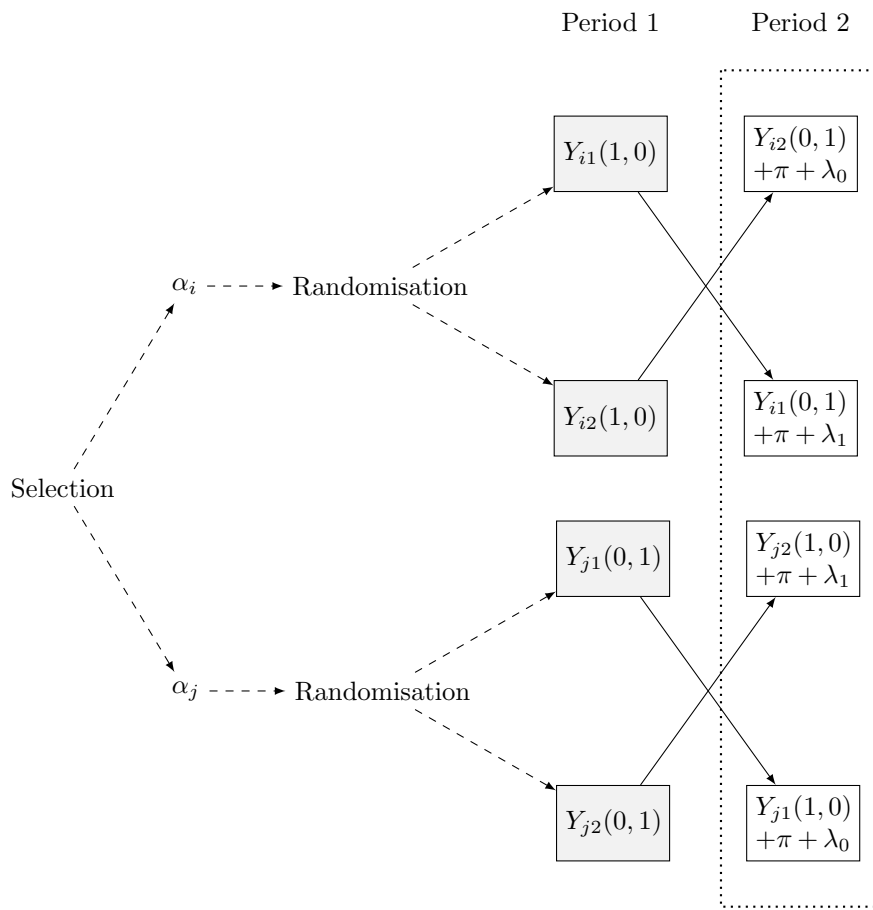

Consider the following within-family ABBA cross-over design. A set of families are selected into the study. They do not necessarily constitute a random sample and indeed the selection may influence the overall outcome level in the family ($\alpha$). For a given family, randomisation is performed to assign each of two siblings to one of two arms, exposed-unexposed or unexposed-exposed, and in the figure we suppose for family $i$ that the outcome of the randomisation is that sibling one is first exposed and then unexposed end \emph{vice versa} for family $j$. There is a potential period effect as represented by $\pi$ and we further assume that the exposure status may carry over from period one to period two, as represented by the carry-over effects $\lambda_0$ and $\lambda_1$. Note the assumption that the cross-over trial may be designed to avoid carry-over of the interference effect. An idealised version of the design is given in the schematic in Figure \ref{fig:schematic-cross-over}.

Recall that we write $Y_{ij}(x_1, x_2)$ for the counterfactual outcome in family $i$ on sibling $j$ when, possibly counterfactually, the first sibling's exposure is set to $x_1$ and the second sibling's exposure is set to $x_2$. Proceeding with the conventional analysis conventional analysis averaging the within-person period differences between arms in this idealised setup yields,
\begin{equation}
  \label{eq:31}
  \begin{aligned}
    \text{Cross-over exposure effect}
    &=
      \frac{1}{2}
      \left(
      \expect{Y_1(1,0) - Y_1(0,1) - \pi - \lambda_1} + \expect{Y_2(0,1) - Y_2(1,0) + \pi + \lambda_0}
      \right) \\
    &=
      \frac{1}{2}
      \left(
      \expect{Y_1(1,0) - Y_1(0,1)} + \expect{Y_2(0,1) - Y_2(1,0)}
      \right)
      + \frac{\lambda_0 - \lambda_1}{2} .
  \end{aligned}
\end{equation}
A common assumption would be that of symmetry (such as considered more formally above) in the sense that the treatment contrasts on sibling two would agree in expectation with the contrast in sibling one, so that the calculation could continue as,
\begin{equation}
  \label{eq:18}
  \begin{aligned}
    \text{Cross-over exposure effect}
    &=
      \expect{Y_1(1,0) - Y_1(0,1)}
      + \frac{\lambda_0 - \lambda_1}{2} \\
    &=
      \frac{1}{2}
      \left(
      \target{2}(1) + \target{2}(0)
      \right) \\
    &\qquad +
      \frac{1}{2}
      \left(
      \left\{
      \lambda_0 - \lambda_1
      \right\}
      -
      \left\{
      \expect{Y_1(1, 1) - Y_1(0, 0) - Y_1(1,0) + Y_1(0,1)}
      \right\}
      \right) ,
  \end{aligned}
\end{equation}
where we have used the same calculations as those leading to equation (\ref{eq:30}). The derivation shows that the difference in carry-over effect is aliased with the deviation from the NDI assumption in the sense that a carry-over effect in the target trial would yield interference in the observational study even if there was no interference effect and conversely, an interference effect would be mistaken as a carry-over in the cross-over trial. Thus, we might say that the terminology from \cite{sjolander_carryover_2016} where interference is referred to as carry-over is warranted. Further, the calculation motivates the NDI assumption as corresponding to the assumption of non-differential carry-over in the hypothetical cross-over target trial. Finally, the derivation shows that the matched-pairs analysis under interference will estimate the same effect as our cross-over design in the presence of differential carry-over. In the absence of a differential carry-over effect, our cross-over design presents an opportunity to estimate the target parameter $\target{3}$.

Note that the difference $\lambda_1 - \lambda_0$ may be estimated by comparing the within-period effect estimates across periods (or, equivalently, comparing the within-sibling sums between arms). From equation (\ref{eq:18}) we may also identify the difference and make the, perhaps natural, identification of $\lambda_1$ with $\expect{Y_1(1, 1) - Y_1(1,0)}$ and $\lambda_0$ with $\expect{Y_1(0, 0) - Y_1(0,1)}$, so that $\lambda_1$ (or $\lambda_0$) is the effect of intervening by removing (adding) exposure in an exposed (unexposed) family.

Naturally, we do not observe the target cross-over trial in its entirety, but may imagine that we are observing only the second period data in the experiment, as indicated in Figure \ref{fig:schematic-cross-over}. Thus, we are left with an exposure discordant data set with an interference effect. To further elucidate the role of randomisation we consider a standard model \citep[e.g.][]{jones2003design} for the cross-over data by introducing an additional subscript and write $Y_{ijp}$ for the observation of the $j$'th sibling from the $i$'th family in period $p$. Let $A_{ij} \in \setdef{1,2}$ be the randomisation arm of sibling $j$ in family $i$. Suppose further that $A_{ij} = 1$ corresponds to the exposure-nonexposure arm, while $A_{ij} = 2$ means nonexposure-exposure. Consider the model,
\begin{equation}
  \label{eq:42}
  Y_{ijp} = \mu + \pi \left[ p - 1 \right] + \beta \left[ \left(2 - A_{ij} \right) \left(2 - p\right) + \left(A_{ij} - 1 \right) \left(p - 1\right) \right] +
  \lambda_{2 - A_{ij}} \left[p - 1 \right] + \alpha_i + s_{ij} + \epsilon_{ijp} 
\end{equation}
The within-sibling period differences as considered above are,
\begin{equation}
  \label{eq:43}
  \begin{aligned}
    \delta_{ij}
    &=
      Y_{ij1} - Y_{ij2} \\
    &=
      (-1)^{3 - A_{ij}} \beta - \pi - \lambda_{2 - A_{ij}} + \left( \epsilon_{ij1} - \epsilon_{ij2} \right) .
  \end{aligned}
\end{equation}
In order to estimate the treatment effect, we again compare the expected period difference between the two arms, where the the expectation is taken conditional on $A_{ij} = 1$ and compared to the expectation where $A_{ij} = 2$. This yields $\beta + (\lambda_0 - \lambda_1)/2$ \emph{if} the error term differences agree in mean between the strata $A_{ij} = 1$ and $A_{ij} = 2$. This latter condition may, however, be expected to hold due to randomisation of the treatment arm $A_{ij}$. This argument may be extended slightly: As already noted, we will in our sibling comparison design only observe data from the second period. A natural question is to which extend the cosibling may stand in for the missing first period observation on the sibling. Consider the period differences substituting the cosibling's observation for the first period observation,
\begin{equation}
  \label{eq:44}
  \begin{aligned}
    \tilde{\delta}_{ij}
    &=
      Y_{i(3 - j)2} - Y_{ij2} \\
    &=
      (-1)^{3 - A_{ij}} \beta +
      (-1)^{3 - A_{ij}} \left[\lambda_{0} - \lambda_{1} \right] +
      \left[s_{ij} - s_{i(3 - j)} \right] +
      \left[ \epsilon_{ij2} - \epsilon_{i(3 - j)2} \right] .
  \end{aligned}
\end{equation}
Averaging the difference between arms (corresponding to comparing the two arms in the second period) yields $\beta + (\lambda_0 - \lambda_1)$ \emph{if} the error terms and sibling-specific effects agree in mean between the arms. Again, this may be argued due to the randomisation. The period effect is immaterial, as we only observe data from the same period. Note that the loss of the first period data has doubled the effect of carry-over. However in the absence of carry-over, letting the cosibling stand in for the sibling's first period yields the same as in the full carry-over design. 

As a final consequence of the within-family randomisation, note that the distribution of $\alpha_i$ will be the same regardless of the arm to which the first sibling is randomised. If the model in (\ref{eq:42}) were structural (i.e. is the true model for the counterfactual outcomes), then we see that exchangeability corresponds to the independence between the randomisation arm and the error term differences in (\ref{eq:44}). Perceiving $\alpha$ as capturing all family-level confounding from $U$ (a notion we will return to below), then we see that the assumption (C1) corresponds exactly to the within-family randomisation.

In observing only the second period from the cross-over design, we are left with an exposure discordant data set with an interference effect. Whether the effect estimated by the design can be given a causal interpretation depends on whether the randomisation of individuals within families holds. If so, we may avoid confounding by comparing the two siblings and effectually let the cosibling's second period observation play the role of the missing first period observation on the sibling. As noted, families may constitute a selected sample from the population, and another concern would ask to what extend the estimated effect in this particular selection extends to the population of interest. In short, characteristics that align with the matched pairs analysis as outlined in the introduction.

However, data from a sibling comparison design is usually not limited to the exposure discordant pairs. If the first period data were observed, such concordant pairs could be used to quantify the carry-over effect (the main idea underlying the so-called Balaam cross-over design, e.g. \cite{lairdAnalysisTwoperiodCrossover1992}). We may thus imagine the data in our sibling comparison design as arising as follows: A group of two-sibling families are selected into the study and some of these are selected into the ABBA sub-study, which yields our exposure discordant pairs when restricting to the second period. Other families are selected into other two armed cross-over designs yielding the exposure concordant pairs from the second period. Even in our imperfect observation scheme, the concordant pairs may still contribute an important piece to the puzzle, since they may be used to estimate the distribution of overall family outcome levels (i.e. $\alpha$). If $\alpha$ is informative for the selection process, it may be used to reverse it, as we explore below using weighting. Note that this may be extended to an argument against using matched pairs estimation methods for the sibling comparison, as these discard the data from concordant exposure pairs. Indeed, the same estimator may be obtained from the between-within model, as we outline below, while retaining this information (also see \cite{carlinRegressionModelsTwin2005} and \cite{seaman_review_2014}). We also note, that in our standard analysis model in (\ref{eq:42}), the family effect $\alpha_i$ is additive with respect to the treatment arm, an assumption we will explore in greater detail below.

\section{Representativity and additivity}
\label{sec:repr-addit}

As seen above, a non-differential carry-over effect may seriously hamper the causal interpretation of a carry-over design as, analogously, the interference effect obstructs the interpretation of the sibling comparison. Another issue pertains to the notion of ``representativity'' -- to which extend is the effect in the population from which individuals are sampled for the experiment relevant to the population of individuals who might receive the intervention? Arguably, the expectations in equations (\ref{eq:31}) and (\ref{eq:18}) relate to the selected population from which families are included into the trial, and not to the population of individuals about whom we may wish to draw inference.

More formally, let $S \in \setdef{0, 1}$ be a variable that indicates if an individual was sampled in the study. The question is then to which extent a target conditional to $S=1$ corresponds to the population target, e.g. for the target $\target{3}$, to which extent does $\cexpect{Y_1(1,1) - Y_1(0,0)}{S = 1}$ agree with $\expect{Y_1(1,1) - Y_1(0,0)}$? If data in the study is sampled at random from the population, it might be reasonable to assume that the target in $S=1$ equals the target in the intended population. However, this may not be case for example in a cross-over study. Such a study might be used for instance to establish bioequivalence \citep{sennStatisticalIssuesBioequivalance2001} between a standard and a novel formulation of a drug. Such studies are routinely performed in healthy, younger volunteers, although the drug is intended as a therapeutic in a specific group of patients. Obviously, in such cases additional considerations are needed to extrapolate the effect in the study to the intended population. An important point is that representativity here is not concerned with the representativity of the study participants, but the representativity of the effect \citep{sennGraphicalRepresentationClinical1990} -- while the participants may not be ``representative'' we might still argue that the within-participant effect in the study could represent the same effect in a member of the intended population. Returning to the within-family cross-over design described above, we might imagine that the families are not sampled randomly but according to variables that in turn affect the family's overall level of outcome, say $\alpha$. In this case we might still argue that though the families are a selected subset of the families about whom we wish to draw inference, we can still view the within-family effect in the sample as representative for the within-family effect in the population. More formally, we might have $\cexpect{Y_1(1,1) - Y_1(0,0)}{S = 1, \alpha} = \cexpect{Y_1(1,1) - Y_1(0,0)}{\alpha}$. Note that this does not give a direct way to compute $\target{3}$ since we will then need to integrate these conditional effects with respect to the distribution of $\alpha$ in the population, i.e. to ``transport'' \citep[e.g.][]{elliottImprovingTransportabilityRandomized2023} the conditional estimate to the intended population, and this distribution may not be known. A key argument however is that this transportation is only necessary if $\alpha$ (or the variables influencing $\alpha$) are effect modifiers for the exposure-outcome relation. In other words, if the effect of the family-level variables are additive with respect to the exposures, we may interpret the conditional effect $\cexpect{Y_1(1,1) - Y_1(0,0)}{\alpha}$ as the effect in the population. Otherwise, we will need to account for the fact that the distribution of the effect modifiers is different in the sample compared to the target population.

Returning from the discussion of clinical trials to observational data, we may, using the conventional epidemiological terminology, say that the stratum $S=1$ defines the source population, while $S=S$ is the target population. A specific realisation of $(X_1, X_2, Y_1, Y_2)$ on the event $\setdef{S=1}$ corresponds to an observed data point. As an example, consider a study aiming to quantify the effect of the exposure among the target population composed of all sibling pairs from a source population of twins, so that $S=1$ corresponds to being a twin pair. If it is thought that control for genetic factors is important to argue conditional exchangeability we might better achieve this by comparing twins. The question is then to what extend we may generalise the effect to all sibling pairs. In applications where the abscence of an interference effect could be justified, we might even seek to generalise to any given individual.

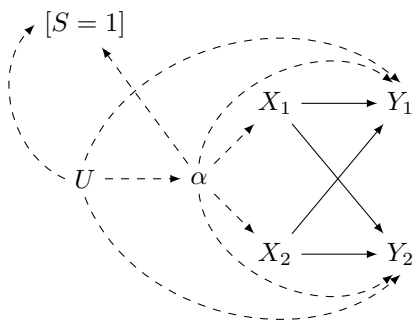
\begin{figure}[htb]
  \centering
  \begin{tikzpicture}
    \node[] (U) {$U$};
    \node[right = of U] (a) {$\alpha$};
    \node[above = 1.5cm of U] (S) {$\left[S = 1\right]$};
    \node[right = of a, above of = a] (X1) {$X_1$};
    \node[right = of a, below of = a] (X2) {$X_2$};
    \node[right = of X1] (Y1) {$Y_1$};
    \node[right = of X2] (Y2) {$Y_2$};
    \draw[->] (X1) -> (Y1);
    \draw[->] (X2) -> (Y2);
    \draw[->] (X1) -> (Y2);
    \draw[->] (X2) -> (Y1);
    \draw[->, dashed] (U) -- (a);
    \draw[->, dashed] (a) -- (X1);
    \draw[->, dashed] (a) -- (X2);
    \draw[dashed] (a.90) edge[bend left = 60, ->] (Y1.110);
    \draw[dashed] (a.270) edge[bend right = 60, ->] (Y2.250);
    \draw[dashed] (a) edge[->] (S);
    \draw[dashed] (U.90) edge[bend left = 55, ->] (Y1.90);
    \draw[dashed] (U.270) edge[bend right = 55, ->] (Y2.270);
    \draw[dashed] (U.180) edge[bend left = 55, ->] (S.180);
  \end{tikzpicture}
  \caption{DAG representing a potential expansion of the setup in
    Figure \ref{fig:dag-assumed}.}
  \label{fig:dag-potential}
\end{figure}

We will now supply the details for this argument in the sibling comparison. Consider Figure \ref{fig:dag-potential}, an expansion of the DAG in Figure \ref{fig:dag-assumed}. The DAG now includes the selection indicator $S$ described above with the square brackets indicating that we are in the subpopulation defined by selection. Further, we have split the unmeasured confounders into $U$ and $\alpha$ corresponding jointly to the variable $U$ in Figure \ref{fig:dag-assumed}. Both $U$ and $\alpha$ are allowed to contribute to the probability of being selected.

We work under the assumptions from Section \ref{sec:setup-notation} but constrain these assumptions to only apply to the selected subpopulation (i.e. all statements are made conditional to the event $S=1$). We consider the following model structure,
\begin{equation}
  \label{eq:11}
  \tilde{M}:\quad
  Y_{ij} = \mu + \alpha_i + \beta^w X_{ij} + \epsilon_{ij} , \quad
  \text{for } j = 1,2, \text{ and } i = 1, \ldots, N ,
\end{equation}
thereby including the variable $\alpha$ from Figure \ref{fig:dag-potential} as a (centered) family-level random intercept.

In Figure \ref{fig:dag-potential}, the variable $\alpha$ may be viewed as a part of $U$ in the original setup in Figure \ref{fig:dag-assumed}. Thus, $\alpha$ is a common cause of the exposures but may also affect the outcomes (note that for this reason, $\alpha$ is not an instrumental variable), and is thus a confounder for the exposure-outcome relationship. From (\ref{eq:11}), $\alpha$ varies only on the family level (which is not easily represented in a DAG), and combining these two observations, we could think of $\alpha$ as representing family-constant confounding.

Although it may notationally appear to be the case, $\tilde{M}$ in (\ref{eq:11}) is not a mixed model. For this to hold, we would need the assumption that $\alpha$ were independent of $\left(X_1, X_2 \right)$, which is clearly at odds with the assumed relationship in Figure \ref{fig:dag-potential} \citep[a point made for instance by][]{sjolander_carryover_2016, brumback_use_2017}, where $\alpha$ is a confounder for the exposure-outcome relationship and thus must be associated with the exposures. Consequently, the approach needs to be modified slightly and we do so by assuming that the family-specific intercept in model $\tilde{M}$ is assumed to depend linearly on the exposures,
\begin{equation}
  \label{eq:26}
  \alpha_i = g \left(X_{i1}, X_{i2} \right) + \tilde{\alpha}_i ,
\end{equation}
for some function $g$. In this modified approach we may more realistically assume that $\tilde{\alpha} \sim N \left(0, \omega^2 \right)$ is independent of $\left(X_1, X_2 \right)$, since the dependence is now contained in the family-specific intercept $\alpha$, and plugging (\ref{eq:26}) into (\ref{eq:11}) we obtain a mixed model. When $g(X_1, X_2) = \left[\beta^B - \beta^W \right] \bar{X}_.$ for a parameter $\beta^B$, the model is a random intercept version of the between-within model \citep{neuhaus_between-_1998}.

The maximum-likelihood estimator of $\beta^w$ in the mixed model $\tilde{M}$ is the GLS estimator. \citeauthor{seaman_review_2014} show that the GLS estimator in the random intercept between-within model coincides with the conditional estimator \citep[][Appendix C]{seaman_review_2014}. We can therefore reuse the result of \citeauthor{petersen2020causal} to establish that,
\begin{equation}
  \label{eq:13}
  \hat{\beta}^w_{\text{GLS}} \longrightarrow
  \cexpect{Y_1}{X_1 = 1, X_2 = 0, S=1} - \cexpect{Y_1}{X_1 = 0, X_2 = 1, S=1} ,
\end{equation}
where we have now made it explicit that the result is conditional to the selected study population.

The role of the latent variables $\alpha$ and $\tilde{\alpha}$ in the causal setup still needs to be clarified and we modify the assumption of conditional exchangeability (C1) with the following,
\begin{itemize}
\item[($\widetilde{\text{C1}}$)] A set of, possibly unmeasured, variables $U$ are sufficient to control for confounding along with $\alpha$ so that\\ \mbox{$Y_j(x_1, x_2) \indep \left(X_1, X_2\right) \mid U, \alpha, S=1$} for both $j=1,2$. Further assume that,
    \begin{equation}
      \label{eq:21}
      \cproba{\alpha \in \cdot}{X_1=x_1, X_2=x_2, S=1} = \cproba{\alpha \in \cdot}{X_1=x_2, X_2=x_1, S=1}
    \end{equation}
  for all possible values $x_1, x_2$.
\end{itemize}
Thinking of $\alpha$ as a part of the unmeasured confounders from the setup in Figure \ref{fig:dag-assumed}, it seems natural to continue to require conditional exchangeability (the first part of ($\widetilde{\text{C1}}$)) while we have only required the symmetric confounding (second part of $(\widetilde{\text{C1}})$) to hold for $\alpha$. We have also made it explicit that these assumptions only need to hold in the source population $S=1$. We retain the consistency assumption (C2). Finally, note the relationship between the second part of assumption $(\widetilde{\text{C1}})$ and the within-family randomisation in the cross-over design in Figure \ref{fig:schematic-cross-over}.

As discussed above, we will assume that the exposure discordant within-family effect in the study population is representative of the effect in the target population, as contained in the following assumption,
\begin{itemize}
\item[($\text{R}$)] Representativity of the family-specific effect holds:
  \begin{equation}
    \label{eq:8}
    \cexpect{Y_1(1, 0)- Y_1(0, 1)}{S = 1, \alpha = a} = \cexpect{Y_1(1, 0)- Y_1(0, 1)}{\alpha = a}
  \end{equation}
\end{itemize}
We will consider a final assumption,
\begin{itemize}
\item[($\text{SA}$)] Structural additivity is said to hold if the causal contrast $Y_1(x, 1 - x) - Y_1(1 -x, x)$ is independent of $\alpha$ for $x = 0, 1$. We say that (SA) holds in mean if,
  \begin{equation}
    \label{eq:22}
    \cexpect{Y_1(x, 1 - x) - Y_1(1 - x, x)}{\alpha} = \expect{Y_1(x, 1 - x) - Y_1(1 - x, x)} ,
  \end{equation}
  for $x = 0, 1$.
\end{itemize}
The (SA) assumption states that we may eliminate the family-specific intercept in exposure-discordant comparisons for the counterfactual outcomes in the same fashion as in the model for the observed outcomes. In other words, the family level has an additive effect on the exposure discordant families, i.e. it is not an effect modifier for the exposure discordant effect. Note that the statistical assumptions (S1) and (S2) and the causal assumptions $(\widetilde{\text{C1}})$ and (C2) need to hold in the study population $S=1$, while the no effect modification assumptions (SA) and (NDI) are assumptions concerning the entire target population.

The argument may then proceed as follows, where we explicitly go through the steps in the derivation to highlight the importance of the various assumptions. Here LTE is the Law of Total Expectation and we represent by $(\ast)$ the observation that \mbox{$U \indep (X_1,X_2) \mid \alpha, S = 1$} in the DAG in Figure \ref{fig:dag-potential}. Thus,

\newpage 

\begin{equation}
  \label{eq:12}
  \begin{aligned}
    &\cexpect{Y_1}{X_1 = 1, X_2 = 0, S = 1} - \cexpect{Y_1}{X_1 = 0, X_2 = 1, S = 1} \\
    &\overset{(C2)}{=}
      \cexpect{Y_1(1, 0)}{X_1 = 1, X_2 = 0, S = 1} - \cexpect{Y_1(0,1)}{X_1 = 0, X_2 = 1, S = 1} \\
    &\overset{(LTE)}{=}
      \int
      \left( \vphantom{\int} \right. \\
    &\qquad\qquad\int
      \cexpect{Y_1(1, 0)}{X_1 = 1, X_2 = 0, S = 1, \alpha = a, U = u}
      f_{U \mid \alpha, X_1, X_2, S} (u \mid a,1, 0, 1)
      \: du \\
      &\qquad\left. \vphantom{\int} \right)
      f_{\alpha \mid X_1, X_2, S} (a \mid 1, 0, 1)
      \: da \\
    &\quad -
      \int
      \left( \vphantom{\int} \right. \\
    &\qquad\qquad\int
      \cexpect{Y_1(0, 1)}{X_1 = 0, X_2 = 1, S = 1, \alpha = a, U = u}
      f_{U \mid \alpha, X_1, X_2, S} (u \mid a, 0, 1, 1)
      \: du \\
      &\qquad\left. \vphantom{\int} \right)
      f_{\alpha \mid X_1, X_2, S} (a \mid 0, 1, 1)
      \: da \\
    &\overset{(\ast, \widetilde{C1})}{=}
      \int
      \left( \int
      \cexpect{Y_1(1, 0)}{S = 1, \alpha = a, U = u}
      f_{U \mid \alpha, S} (u \mid a, 1)
      \: du \right)
      f_{\alpha \mid X_1, X_2, S} (a \mid 1, 0, 1)
      \: da \\
      &\qquad -
      \int
      \left( \int
      \cexpect{Y_1(0, 1)}{S = 1, \alpha = a, U = u}
      f_{U \mid \alpha, S} (u \mid a, 1)
      \: du \right)
      f_{\alpha \mid X_1, X_2, S} (a \mid 0, 1, 1)
      \: da \\
    &\overset{(\widetilde{C1})}{=}
      \int
      \left( \int
      \cexpect{Y_1(1, 0)}{S = 1, \alpha = a, U = u}
      f_{U \mid \alpha, S} (u \mid a, 1)
      \: du \right)
      f_{\alpha \mid X_1, X_2, S} (a \mid 1, 0, 1)
      \: da \\
      &\qquad -
      \int
      \left( \int
      \cexpect{Y_1(0, 1)}{S = 1, \alpha = a, U = u}
      f_{U \mid \alpha, S} (u \mid a, 1)
      \: du \right)
      f_{\alpha \mid X_1, X_2, S} (a \mid 1, 0, 1)
      \: da \\
    &\overset{(LTE)}{=}
      \int
      \cexpect{Y_1(1, 0)- Y_1(0, 1)}{S = 1, \alpha = a}
      f_{\alpha \mid X_1, X_2, S} (a \mid 1, 0, 1)
      \: da \\
    &\overset{(R), (\widetilde{C1})}{=}
      \int
      \cexpect{Y_1(1, 0) - Y_1(0, 1)}{\alpha = a}
      f_{\alpha \mid X_1, X_2, S} (a \mid x, 1-x, 1)
      \: da \qquad\qquad\qquad (\text{\Large\dag}) \\
    &\overset{(R)}{=}
      \int
      \cexpect{Y_1(1, 0) - Y_1(0, 1)}{\alpha = a, S = 1}
      f_{\alpha \mid X_1, X_2, S} (a \mid x, 1-x, 1)
      \: da \\
    &\overset{(LTE)}{=}
      \int
      \left(
      \int
      \cexpect{Y_1(1, 0) - Y_1(0, 1)}{\alpha = a, S = 1, U = u} f_{U \mid \alpha, S} (u \mid a, 1) \: du
      \right)
      f_{\alpha \mid X_1, X_2, S} (a \mid x, 1-x, 1)
      \: da \\
    &\overset{(\widetilde{C1}), (*)}{=}
      \int
      \left(
      \int
      \cexpect{Y_1(1, 0) - Y_1(0, 1)}{\alpha = a, S = 1, U = u, X_1 = x, X_2 = 1-x} f_{U \mid \alpha, S, X_1, X_2} (u \mid a, 1, x, 1-x) \: du
      \right) \\
    &\qquad\qquad\qquad
      f_{\alpha \mid X_1, X_2, S} (a \mid x, 1-x, 1)
      \: da \\
    &\overset{(LTE)}{=}
      \int
      \cexpect{Y_1(1, 0) - Y_1(0, 1)}{\alpha = a, S = 1, X_1 = x, X_2 = 1-x} 
      f_{\alpha \mid X_1, X_2, S} (a \mid x, 1-x, 1)
      \: da \\
    &\overset{(LTE)}{=}
      \cexpect{Y_1(1, 0) - Y_1(0, 1)}{S = 1, X_1 = x, X_2 = 1 - x} \\
  \end{aligned}
\end{equation}
for both $x=0$ and $x = 1$. This reproduces the result from \cite{petersen2020causal}, here explicitly stating that the effect is conditional to the study population. If we, continuing the example above, sought to quantify the effect among sibling pairs from twins, the arrived at effect would pertain only to twins. If the assumption of structural additivity is warranted, we see by picking up from the obelus in equation (\ref{eq:12}),

\begin{equation}
  \label{eq:14}
  \begin{aligned}
    &\cexpect{Y_1}{X_1 = 1, X_2 = 0, S = 1} - \cexpect{Y_1}{X_1 = 0, X_2 = 1, S = 1} \\
    &=
      \int
      \cexpect{Y_1(1, 0) - Y_1(0, 1)}{\alpha = a}
      f_{\alpha \mid X_1, X_2, S} (a \mid x, 1-x, 1)
      \: da \\
    &\overset{(SA)}{=}
      \expect{Y_1(1, 0) - Y_1(0, 1)}
      \int
      f_{\alpha \mid X_1, X_2, S} (a \mid x, 1-x, 1)
      \: da \\
    &=
      \expect{Y_1(1, 0) - Y_1(0, 1)} \\
    &\overset{(NDI)}{=}
      \expect{Y_1(1, 1) - Y_1(0, 0)} 
  \end{aligned}
\end{equation}
so that we obtain an effect that does not suffer from being conditional to neither the exposure discordant population, nor the study population. If, moreover, non-differential interference holds, we obtain the target parameter $\target{3}$.

Aside from the noted connection to the cross-over design, there is also an obvious analogue between the matched analysis in the sibling comparison to a more classic matched design, as outlined in Appendix \ref{sec:append-match-comp}.

\subsection{Reversing selection by weighting}
\label{sec:revers-select-weight}

As shown in the preceding section there is a discrepancy between the target of the matched analysis and those that would usually be targeted by a sibling comparison design. The discrepancy may be summarised as the fact that the target of the matched analysis is
\begin{equation}
  \label{eq:45}
  \cexpect{Y_1(1, 0) - Y_1(0, 1)}{S = 1, X_1 = x, X_2 = 1 - x} ,
\end{equation}
as shown in equation (\ref{eq:12}) above. As described in Section \ref{sec:target-trials} we may think of the conditioning event as arising from two stages of selection, selection into the source population ($S=1$) and selection into the ABBA cross-over study ($X_1 = x, X_2 = 1-x$). We need the (SA) assumption to hold for these selections to be ignored. In order to estimate one of our previously defined targets, we further need (NDI) to hold, and in this case, the matched analysis estimates the target parameter $\target{3}$.

An alternative strategy to counter the effects of selection bias is to assign different weights to observations in the analysis to account for their potentially differing representativeness. Thus, we would assign larger weights to observations that are less likely to be observed in the data the idea being that if this weighting is proportional to the frequency of such observations in the target population we can thereby restore the representativity of the data. We discuss this weighting strategy in two stages of increasing difficulty. First, we discuss weighting within the study to restore representativity across sub-studies in order to obtain an, ideally, unbiased estimate of a conditional target given $S=1$, e.g. $\cexpect{Y_1(1,1) - Y_1(0,0)}{S=1}$, a conditional version of $\target{3}$. If we think of $S=1$ as being twins, we are thus looking to obtain an estimate of, say, $\target{3}$ for twins that does not condition on exposure discordance. We return to the more difficult problem of reversing the other stage of selection in the discussion.

Consider the weight given by,
\begin{equation}
  \label{eq:35}
  w(a, x_1, x_2) = \indic{X_1 = x_1, X_2 = x_2} \left[ f_{(X_1, X_2) \mid \alpha, S} (x_1, x_2 \mid a, 1) \right]^{-1}
\end{equation}
We intend to use weighted observations of the form $Y_{ij} w(\alpha_i, x_1. x_2)$ to estimate counterfactual means. The argument may be summarised as follows. Consider sibling $j$ in a given family (we suppress the family subscript), so that,
\begin{equation}
  \label{eq:33}
  \begin{aligned}
    &\cexpect{Y_{j} w(\alpha, x_1, x_2)}{S = 1}
    \\
    &=
      \int
      \left[ f_{(X_1, X_2) \mid \alpha, S} (x_1, x_2 \mid a, 1) \right]^{-1}
      \cexpect{Y_{j} \indic{X_1 = x_1, X_2 = x_2}}{S = 1, \alpha = a, U = u}
      \\
      &\quad\qquad \cdot f_{(\alpha, U) \mid S} (a, u \mid 1)
      \: (da, du)
    \\
    &=
      \int
      \frac{f_{(X_1, X_2) \mid \alpha, U, S} (x_1, x_2 \mid a, u, 1)}{f_{(X_1, X_2) \mid \alpha, S} (x_1, x_2 \mid a, 1)}
      \cexpect{Y_{j} (x_1, x_2) }{X_1 = x_1, X_2 = x_2, S = 1, \alpha = a, U = u}
      \\
      &\quad\qquad \cdot f_{(\alpha, U) \mid S} (a, u \mid 1)
      \: (da, du)
      \\
    &=
      \int
      1 \cdot
      \cexpect{Y_{j} (x_1, x_2) }{S = 1, \alpha = a, U = u}
      f_{(\alpha, U) \mid S} (a, u \mid 1)
      \: (da, du)
      \\
    &=
      \cexpect{Y_{j} (x_1, x_2) }{S = 1}
  \end{aligned}
\end{equation}
The first equality marginalises over confounders. The second uses consistency and conditional exchangeability. Noticing again that $(X_1, X_2) \indep U \mid \alpha$ yields the third equality, and we reverse the marginalisation in the final equality.

An obvious strenght of this approach is that it does not rely on neither the structural additivity (SA) nor the non-differential interference (NDI) assumption. Further, it may be used to estimate any given target within the selection. There are two main downsides of the method, both of which are practical. First, we do not know the conditional probability of exposure given confounders ($f_{(X_1, X_2) \mid \alpha, S} (x_1, x_2 \mid a, 1)$), something that is a recurring challenge for weighting methods. The second issue is salient to the application to sibling comparisons: In practice we do not know the confounder $\alpha$. The former issue may be approached through a model for the exposure assignment as described in the weighting literature \citep[as reviewed in][]{seamanReviewInverseProbability2011}. We propose to tackle the latter problem in the same way and replace $\alpha$ by an estimate $\hat{\alpha}$. One could for example fit the between-within model from equation (\ref{eq:11}) and (\ref{eq:26}) and use,
\begin{equation}
  \label{eq:37}
  \hat{\alpha}_i = \bar{X}_{i .} \left( \hat{\beta}^B - \hat{\beta}^W \right) + \widehat{\tilde{\alpha}}_i ,
\end{equation}
as an estimate for $\alpha$, where $\widehat{\tilde{\alpha}}$ is the best linear unbiased predictor \citep[BLUP,][]{demidenko_mixed_2013} for the $i$'th family. Naturally, other models could form the basis for estimation of $\alpha$, particularly if some of the variables in $U$ were observed.

In the following, we explore the performance of this method in a small simulation study before applying it to a data example. 

\section{Simulation study}
\label{sec:simulation-study}

We simulate data from the model,

\begin{equation}
  \label{eq:39}
  Y_{ij} = \mu + \alpha_i + \beta^W \cdot X_{i j} + \beta^D \cdot X_{i j} \cdot \alpha_i + \beta^C \cdot X_{i (3 - j)} + \epsilon ,
\end{equation}
where,
\begin{equation}
  \label{eq:41}
  \alpha = b_{\alpha} \cdot \exp(b_{\Lambda} \cdot U) + \tilde{\alpha},
\end{equation}
for $\tilde{\alpha} \sim N(0, \tau^2)$. The distribution of $X_{ij}$ depends on $U \sim N(0, \sigma_U^2)$, which is a confounder. The details of the simulations are given in Appendix \ref{cha:append-deta-simul}. As expanded on in the same appendix, the parameters $\beta^D$ and $\beta^C$ represent deviations from (SA) and (NDI), respectively. 

We simulate data under four scenarios, so that neither, one of, or both of the assumptions of structural additivity (SA) and non-differential interference (NDI) are satisfied. For each scenario, 2000 data sets are simulated. For each replication, we analyse the resulting data set using the random intercept between-within (BW) model discussed in Section \ref{sec:repr-addit} and the weighting method discussed in Section \ref{sec:revers-select-weight}. For the weighting method, three different weights are used: The true weight, the estimated weight evaluated at the true $\alpha$, and finally an estimated weight evaluated at an estimated $\alpha$. Logistic regression is applied to estimate the weight, while the estimate of $\alpha$ comes from the BW model as discussed in Section \ref{sec:revers-select-weight}. The two methods are viewed as targeting the conditional version of $\target{3}$ given selection, i.e. \mbox{$\cexpect{Y_1(1, 0) - Y_1(0, 1)}{S = 1, X_1 = x, X_2 = 1 - x}$}. Also note that no attempt is made to reduce the variance of the weighting method by stabilising, truncating, or excluding weights \citep{seamanReviewInverseProbability2011}, albeit such measures may be sensibly taken in applications. When a model did not converge or returned inverse weights that were numerically zero, the procedure returned a missing value.

Table \ref{tab:simulations-summaries} gives the results for the simulations when using the true weights, estimated weights evaluated at the true $\alpha$, and when using estimated weights evaluated at estimated $\alpha$. As predicted from Section \ref{sec:revers-select-weight}, the weighting procedure yields an unbiased estimate of the conditional target $\target{3}$ given $S = 1$ when using the true weights. This is not, however, the case for the BW model outside Scenario 4 where both (SA) and (NDI) hold, as predicted in Section \ref{sec:repr-addit}. In all cases, the standard error of the weighted procedure is considerably higher than in the BW model and in a few cases the weighting procedure produces estimates that are quite widely removed from the true value (presumeably due to extreme weights). In the case of both (SA) and (NDI) (i.e. Scenario 4), the BW model is clearly preferable. When evaluating the estimated weight at the true $\alpha$, the procedure yields unbiased estimated, which is expected because the applied weight model contains the true weight model (see Appendix \ref{cha:append-deta-simul}). Substituting for $\alpha$ an estimate, the weighting procedure exhibits (numerically) larger bias than the BW model in the first three scenarios. In Scenario 4, both estimation procedures are unbiased (the BW model being superior in terms of standard error). We also note that in the first two scenarios, there were some problems with obtaining an estimate (due to numerically zero inverse weights).

\begin{table}[htb]
  \centering
  Scenario 1
\begin{tabular}{lp{2.5cm}p{2.5cm}p{2.5cm}p{2.5cm}}
  \hline
 & BW model & Weighting $w(\alpha)$ & Weighting $\widehat{w}(\alpha)$ & Weighting $\widehat{w}(\hat{\alpha})$ \\ 
  \hline
N & 2000 & 2000 & 2000 & 1977 \\ 
  Mean & 2.42 & 5.85 & 5.86 & 0.53 \\ 
  Bias & -3.42 & 0.01 & 0.02 & -5.31 \\ 
  SE & 0.11 & 0.18 & 0.11 & 2.01 \\ 
   \hline
\end{tabular}

  \\
  Scenario 2
\begin{tabular}{lp{2.5cm}p{2.5cm}p{2.5cm}p{2.5cm}}
  \hline
  \hline
N & 2000 & 2000 & 2000 & 1999 \\ 
  Mean & 3.50 & 6.50 & 6.50 & 3.18 \\ 
  Bias & -3.00 & 0.00 & 0.00 & -3.32 \\ 
  SE & 0.06 & 0.25 & 0.11 & 1.02 \\ 
   \hline
\end{tabular}

  \\
  Scenario 3
\begin{tabular}{lp{2.5cm}p{2.5cm}p{2.5cm}p{2.5cm}}
  \hline
  \hline
N & 2000 & 2000 & 2000 & 2000 \\ 
  Mean & 3.92 & 4.35 & 4.36 & 1.73 \\ 
  Bias & -0.42 & 0.01 & 0.02 & -2.61 \\ 
  SE & 0.11 & 0.18 & 0.11 & 1.31 \\ 
   \hline
\end{tabular}

  \\
  Scenario 4
\begin{tabular}{lp{2.5cm}p{2.5cm}p{2.5cm}p{2.5cm}}
  \hline
  \hline
N & 2000 & 2000 & 2000 & 2000 \\ 
  Mean & 5.00 & 5.00 & 5.00 & 5.00 \\ 
  Bias & -0.00 & 0.00 & 0.00 & -0.00 \\ 
  SE & 0.06 & 0.25 & 0.11 & 0.11 \\ 
   \hline
\end{tabular}

  \caption{Simulation results for the four estimation procedures (columns), the first is the between-within model, the second, third and fourth is the weighting procedure using true weights, estimated weights at true $\alpha$, and estimated weights at estimated $\alpha$, respectively. In the four scenarios (row blocks), we give a number of summary statistics (rows): the number of non-missing estimates (N), their mean, bias, and standard error (SE).}
  \label{tab:simulations-summaries}
\end{table}

\begin{figure}[htb]
  \centering
  \includegraphics[width = \linewidth]{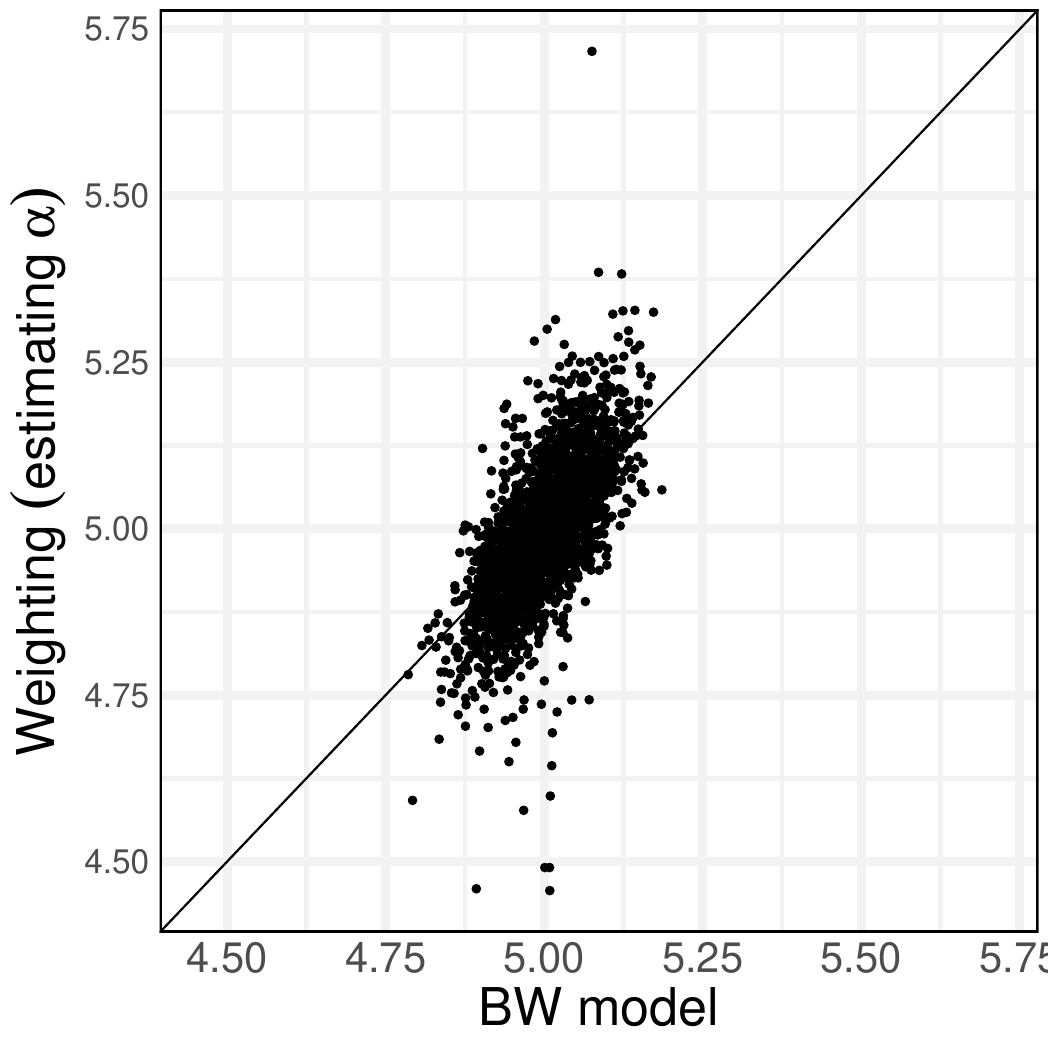}
  \caption{Plot of estimates for $\target{3}$ in Scenario 4. The horisontal axis is the estimates from the BW model, while the vertical axis shows estimates from the weighting procedure evaluating the estimated weight at estimated $\alpha$.}
  \label{fig:simul-BW-and-Weighting}
\end{figure}

Numerical experimentation indicates that the bias may be both larger and smaller for the weighting method compared to the BW model depending on the choice of simulation parameters. The simulations indicate, as is the case more generally with inverse probability weighting, that the procedure depends on a correct weight model. In the present case, this may partitioned into a model for $\alpha$ and the conventional model for the weights given $\alpha$. The simulations show that even when the correct model is contained in the weight model, the procedure is dependent on proper estimates of $\alpha$. Of course this raises the question, why one should bother with the weighting method in the first place. We argue, that it has at least two uses. First, it may be used to assess the assumptions of (SA) and (NDI) for the BW model matched analysis. The argument is one of contradiction: If one believes the BW model to ``hold'', so that (SA) and (NDI) is satisfied, then one would expect it to yield good estimates of $\alpha$, and then the estimates from the BW model and weighted analysis are expected to agree. Thus, one may take discrepant estimates between the two procedures to indiate a violation of the (SA) and (NDI) assumptions. Figure \ref{fig:simul-BW-and-Weighting} shows estimates from the BW model plotted agains those from the weighting method using estimated $\alpha$ in Scenario 4 (where the BW model holds). Disregarding a few extreme estimates from the weighting method, presumeably due to large weights, there is a high agreement between estimates from the two methods (the variance of the weighting procedure being obviously larger). The second use of the method is when one expects the BW model to hold, so that it can supply proper estimates of $\alpha$, since in this case the weighting procedure can target other causal target parameters. We illustrate the first use on an example data set. 

\section{Example: Familial conflict and quality of life}
\label{sec:example:familial-}

We illustrate the proposed methods using data from a study on twins. In \cite{aaModerationGeneticFactors2010}, the authors study the effect of family function on general quality of life and the role of divorce as a potential effect modifier of this relationship using self-reported data on 6,773 twins and siblings. A variety of psychometric scales are collected, including a questionaire on familial conflict (FC) and general quality of life (QoL). The FC consists, as explained in greater detail by \citeauthor{aaModerationGeneticFactors2010}, of ten items scored 1 for ``No'' and 2 for ``Yes'' with each affirmative answer indicating a higher level of conflict in the family. The resulting scores are summed to yield a score between 10 and 20. QoL is measured by the ten step Cantril Ladder, with 10 and 1 representing respectively the best and worst possible life. The dataset is available in the \textsf{R}-package \textsf{umx} (version \textsf{4.21.0} on CRAN).

In our analysis we consider only families of complete twin cases in the sense that both twins in the family must have non-missing data on FC and QoL. FC has been coded as negative in the data as to range from -20 to -10 in the data (presumeably to better agree with the hypothesised direction of correlation of other variables in the data). We let $Y_j$ be the QoL of sibling $j$ and define $X_j = \indic{FC_j < - 15}$ to be the exposure, so that a sibling is considered exposed to familial conflict if they answered affirmative to more than five of the ten questions. Some basic descriptives of the considered sample is given in Table \ref{tab:example-summaries}.

\begin{table}[htb]
  \centering
  \begin{minipage}{0.6\linewidth}
    \fontsize{12.0pt}{14.4pt}\selectfont
\begin{tabular*}{\linewidth}{@{\extracolsep{\fill}}lc}
\toprule
\textbf{Variable} & \textbf{N = 2,515}\textsuperscript{\textit{1}} \\ 
\midrule\addlinespace[2.5pt]
Age & 16.06 (1.60) \\ 
Divorce & 293 (12\%) \\ 
Zygosity &  \\ 
    MZ & 1,000 (40\%) \\ 
    DZ & 1,515 (60\%) \\ 
Sex, Twin 1 &  \\ 
    male & 1,128 (45\%) \\ 
    female & 1,387 (55\%) \\ 
Sex, Twin 2 &  \\ 
    male & 1,102 (44\%) \\ 
    female & 1,413 (56\%) \\ 
X1 &  \\ 
    0 & 1,821 (72\%) \\ 
    1 & 694 (28\%) \\ 
X2 &  \\ 
    0 & 1,802 (72\%) \\ 
    1 & 713 (28\%) \\ 
Xbar &  \\ 
    0 & 1,520 (60\%) \\ 
    0.5 & 583 (23\%) \\ 
    1 & 412 (16\%) \\ 
Y1 & 7.83 (1.12) \\ 
Y2 & 7.85 (1.10) \\ 
\bottomrule
\end{tabular*}
\begin{minipage}{\linewidth}
\textsuperscript{\textit{1}}Mean (SD); n (\%)\\
\end{minipage}

  \end{minipage}
  \caption{Summary statistics for the twin families with non-missing data on $(X_1, X_2, Y_1, Y_2)$. Continous variables are summarised as mean and standard deviation (SD), while categorical variables are summarised by count and percentage of the total.}
  \label{tab:example-summaries}
\end{table}

We assume that the interrelationship of the variables is as in Figure \ref{fig:dag-potential}. Here it should be stressed that due to the cross-sectional nature of the data, this is tentative at best, and is meant mainly as an illustration. We use the selection indicator to indicate that the study is conducted among twins ($S=1$) but that we would ideally like to generalise its findings to the entire population ($S = S$). We also assume the statistical assumptions (S1) and (S2) from Section \ref{sec:setup-notation}.

Table \ref{tab:example-bwmodel-estimates} gives estimates from the random intercept between-within (BW) model from equation (\ref{eq:11}). We estimate that the average QoL among siblings belonging to a family in which neither sibling experiences familial conflict is 8.02 (95 \% CI: [7.98; 8.06]) points. This QoL is reduced by 0.28 ([0.18 ; 0.39]) points on average when comparing an exposed sibling to an unexposed cosibling within families.

\begin{table}[htb]
  \centering
  
\fontsize{12.0pt}{14.4pt}\selectfont
\begin{tabular*}{\linewidth}{@{\extracolsep{\fill}}lcc}
\toprule
\textbf{Characteristic} & \textbf{Beta} & \textbf{95\% CI} \\ 
\midrule\addlinespace[2.5pt]
(Intercept) & 8.02 & 7.98, 8.06 \\ 
X & -0.283 & -0.388, -0.178 \\ 
Xbar & -0.363 & -0.501, -0.225 \\ 
\bottomrule
\end{tabular*}
\begin{minipage}{\linewidth}
Abbreviation: CI = Confidence Interval\\
\end{minipage}

  \caption{Estimates from the between-within model for the familial conflict data.}
  \label{tab:example-bwmodel-estimates}
\end{table}

If we further impose the causal assumptions (C1)-(C3) we may extend this latter interpretation and cast it as an estimate of $\cexpect{Y_1(1, 0) - Y_1(0,1)}{X_1 \not = X_2, S=1}$, cf. equation (\ref{eq:12}). Thus, we estimate that among twin families that are exposure discordant (twins disagree on the experience of familial conflict), the mean QoL among twins who are exposed to familial conflict, while their cotwin is not, is 0.28 ([0.18 ; 0.39]) points lower than the mean QoL among twins belonging to families, where the experience of familial conflict had been reversed.

We supplement the BW model with a weighted analysis to estimate the conditional version of $\target{3}$ given selection, $\cexpect{Y_1(1, 1) - Y_1(0, 0)}{S=1}$, corresponding to the family level intervention comparing an exposed family to un unexposed. As described in Section \ref{sec:revers-select-weight}, we first estimate $\alpha$ in the sample using the BW model. We fit two logistic regressions models assuming that the risk of both/neither twin being exposed is a linear function of $\alpha$ on log-odds scale. These models are used to estimate the weights. The weights are used to estimate the difference in weighted means which as argued above in Section \ref{sec:revers-select-weight} yields an unbiased estimate of $\cexpect{Y_1(1, 1) - Y_1(0, 0)}{S=1}$ (assuming a correct weight model). In order to obtain confidence intervals we bootstrapped the above procedure. Non-parametric cluster bootstrapping was performed sampling families with replacement and we report Wald confidence intervals using the bootstrap standard error based on 1000 replications.

\begin{figure}[htb]
  \centering
  \includegraphics[width=\linewidth]{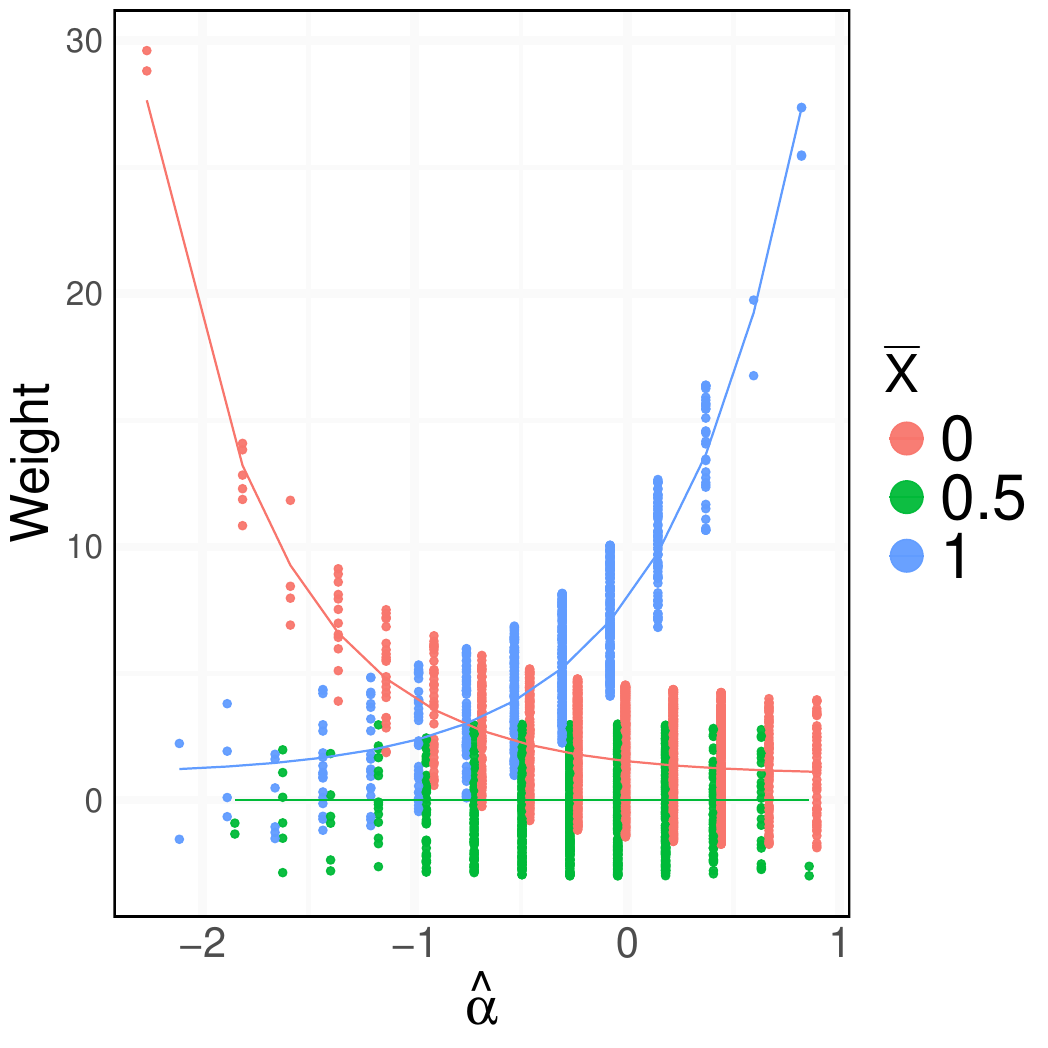}
  \caption{Plot of the weights $\tilde{w}(X_1, X_2, \alpha) = (1-X_1) \cdot (1-X_2) \cdot w(0, 0, \alpha) + X_1 \cdot X_2 \cdot w(1, 1, \alpha)$ with the weight function $w$ as defined in equation (\ref{eq:35}). The weight is evaluated at the estimated $\alpha$ and coloured by the overall family exposure status. Points have been jittered vertically to avoid overplotting and the lines connect the original, unjittered, positions of the points.}
  \label{fig:example-weights}
\end{figure}

The estimated weights are depicted in Figure \ref{fig:example-weights}. We see that there are a few somewhat large weights and as a sensitivity analysis we excluded observations with weights larger than 15 (eight observations corresponding to four families). We estimate that family-level exposure leads to an increase of 0.17 ([0.01 ; 0.33]) QoL points on average among twins. This is a somewhat counterintuitive result, which seems to be at least partly driven by observations with large weights. Excluding the eights observations with largest weights, we estimate that the family-level exposure leads to an increase of 0.05 [-0.08 ; 0.19] QoL points on average. With 95\% confidence we exclude average increases in QoL larger than 0.19 points and exclude decreases larger than 0.08 points. Note that when bootstrapping the sensitivity analysis, we took the decision to remove the largest weights as an instance of the more general rule of excluding weights above the $99.841 \%$ percentile of weights and incorporated this rule into the bootstrap.

In summary, the BW model estimates a small decrease in QoL while the weighted analysis estimates a very small increase, the CI ranging from a very small decrease to a small increase in QoL. The discrepancy between the two analyses may of course be ascribed to the different targets, which differ in two aspects, the first aspect being the difference in the intervention on $(X_1, X_2)$ (comparing $(1, 0)$ to $(0, 1)$ versus $(1,1)$ to $(0,0)$) the second being that the BW model targets a sub-population of exposure discordant families. We discuss the implications of these differences in turn by fixing the other. Consider a scenario where the BW models target is not conditional to the discordant twins (A $S=1$-conditional version of the structural additivity assumption), so that we are effectively comparing $\cexpect{Y_1(1, 1) - Y_1(0, 0)}{S=1}$ to $\cexpect{Y_1(1, 0) - Y_1(0, 1)}{S=1}$. A difference between the two targets would be a sign of non-differential interference. If conversely $Y_1(1, 0) - Y_1(0, 1)$ agrees with $Y_1(1, 1) - Y_1(0, 0)$ on exposure discordant twins (a conditional version of the non-differential interference assumption), the discrepancy simply expresses that the effect when considering all twins is different from the effect found among the exposure discordant. A third, also not unrealistic, explanation of the discrepancy would be a misspecified model for the weights. 

The latter explanation may be at least partially remedied by more elaborate modellering, both in the weight model (however, for example letting the weight depend on $\alpha$ as a quadratic function on log-odds scale did not yield substantially different estimates) and in the model for $\alpha$ for example including measured variables from $U$ (e.g. age, sex and divorce status). We cannot rule out either of the two remaining explanations, but considering the subject matter it seems likely that interference could be present, so that the negative effect of experiencing familial conflict is in a sense countered when the cosibling has the same experience, while the negative impact of conflict on QoL is exacerbated when experiencing familial conflict in a family where the cotwin has the opposite experience. While we cannot reach a definitive explanation, we also note that the discrepancy between the two analyses should at least caution us against assuming either structural additivity or non-differential interference.

\section{Discussion}
\label{sec:discussion}

The key takeaway from the preceeding derivations is: Effect modification. More precisely, the two different types of effect modification that pertain to the NDI and the SA assumptions. These assumptions correspond to the same assumptions needed by the cross-over trialist when arguing non-differential carry-over and representativity of the trial effect, thus accentuating the already established kinship between a twin comparison and the cross-over design \citep{kalow_hypothesis_1998}. This spin on the selection bias seems to be absent from the literature on the shortcomings of the sibling comparison, although it is discussed in these terms in the recent paper by \citeauthor{sjolanderGeneralizabilityEffectMeasure2022} \citep{sjolanderGeneralizabilityEffectMeasure2022} under a somewhat different setup. The representation of the selection bias as depending on effect modification seems more widely appreciated in the literature on matched studies. 

We also remark that the assumption of NDI slightly relaxes the condition of no interference in the strong sense, although as it is also the case for cross-over trials, there is in practice not many examples where it can be reasonably argued that carry-over is present but at the same time non-differential. As such, in most practical applications discussion will center on whether the interference effect is negligible. In the litterature on causal inference, the assumption of no interference is usually subsumed under the so-called SUTVA (stable unit treatment value assignment) assumption \citep{vanderweeleCausalInferenceMultiple2013}. 

Structural additivity may be viewes as analogous to the concept of treatment-by-trial interaction in a clinical trial. The views of treatment-by-trial interaction in randomised studies may be summarised into two schools depending on whether they accept or seek to remedy the situation. An argument of the former is voiced by \citeauthor{petoRepresentationPeople1988} who wrote that ``\emph{qualitative conclusions, and not direct extrapolations of absolute risk reductions from studies of "representative" samples of the population, are what trials can offer}'' \citep{petoRepresentationPeople1988}. Similarly, \citeauthor{sennGraphicalRepresentationClinical1990} writes ``\emph{It is in the hope that trial by treatment interaction is relatively unimportant that clinical trials are carried out at all}.'' \citep{sennGraphicalRepresentationClinical1990}, also see the discussion in \cite{sennCrossoverTrialsDegrees1991}. Proof of existence of the latter school is the body of literature on generalisability and transportability, for example \cite{elliottImprovingTransportabilityRandomized2023}, where the representativity assumption in ($\text{R}$) is cast as the assumption of ``ignorability relevant for transportability'' \citep{elliottImprovingTransportabilityRandomized2023}. Arguably, results from sibling comparisons do not lend themselves well to most methods for transportability, since the variables relevant for arguing transportability are usually unobserved. This is exactly the incentive for employing the design, that we do not need to measure e.g. genetic variables $U$, the downside being that we cannot supply the conditional effects given $U$. Indeed, the structural additivity assumption states that the effect does not depend on $U$ (through $\alpha$).

An interesting line of inquiry to assess the validity of this assumption might be representative probability samples as discussed in  \cite{elliottImprovingTransportabilityRandomized2023} using a sample from the target population. Along these lines it might be worth considering weighting both as a method to obtain exchangeability but also to achieve representative effects. Indeed, consider the weights defined by,
\begin{equation}
  \label{eq:15}
  w(a, x_1, x_2) = \indic{X_1 = x_1, X_2 = x_2, S = 1} \left[f_{(X_1, X_2) \mid \alpha} (x_1, x_2 \mid a)
    f_{S \mid \alpha} (1 \mid a) \right]^{-1}
\end{equation}
corresponding to the product of inverse probabilities from a treatment model $(X_1, X_2) \mid \alpha$ and a selection model $S \mid \alpha$ \citep[also see][Section 12.6]{hernanCausalInferenceWhat2020}. Weighted observations of the form $w(\alpha_i, x_1, x_2) Y_{ij}$ may then be used to unbiasedly estimate $\expect{Y(x_1, x_2)}$ and may thus form the basis for estimation of other targets beside $\target{3}$ under the additional assumption that the probability of selection depends on $U$ only through $\alpha$. The argument is entirely analogous to that in Section \ref{sec:revers-select-weight} and for the sake of completeness we provide it in Appendix \ref{sec:append-weight-proc}. In practise, $\alpha$ will not be observed and may be replaced by the best linear unbiased predictor \citep[BLUP,][]{demidenko_mixed_2013}. Estimation of the treatment and selection model will require a sample from the target population. One could then estimate the BLUPs in this sample and thus estimate the weights. Here one should be mindful that both models would suffer from measurement error dilution due to using the estimated rather than true $\alpha$ \citep{carrollMeasurementErrorNonlinear2006}. The exact manner of prediction could be done in various ways but would require knowledge of both outcomes and exposures in the sample. The crux of the algorithm is the specification of a connection between the distribution of $\alpha$ in the study population and the reference population. One starting point could be to pose a model of the between-within form also in the reference population and use a range of distributions for $\alpha$ as in (\ref{eq:26}) that seem reasonable from background knowledge. Note that we are not interested in attaching any causal interpretation to the exposure effect in this model but simply need it to estimate the confounder distribution. We also stress here that the Gaussian model assumptions on $\alpha$ following equation (\ref{eq:26}) are only used to justify the equality between the fixed effects conditional estimator and the random effects GLS estimator. Other distributional models may indeed also prove relevant. In practise, sufficiently rich external samples will prove difficult to obtain, but one might imagine alternative, more \emph{ad hoc} versions of the procedure. For example, if the exposures along with some of the variables in $U$ are known one might utilise the $U \rightarrow \alpha$ relation to obtain more tentative predictions of $\alpha$ in the external sample and then use these predictions in the treatment and selection models. Since the weighting procedure does not require the NDI or SA assumptions, estimates from such alternative implementations of the procedure may still prove useful as sensitivity analyses for the matched-pair estimator for $\target{3}$. In principle, as showed by the derivation in Appendix \ref{sec:append-weight-proc}, the weighting procedure may be used to estimate any of the targets considered above. However in practice, the necessary model assumptions on the reference population needed to estimate $\alpha$ may be deemed too strong to be considered realistic. Rather, the procedure may be viewed as an opportunity to estimate $\target{3}$ without assuming (SA) and (NDI) and thus to gauge the estimations procedure's sensitivity to these assumptions. Of course, these assumptions of no effect modification would themselves be replaced by assumptions on the confounder distribution. In many scenarios it may, arguably, be easier to define a plausible range of such distributions to assess their influence on the estimate than to assess a plausible range of the two types of effect modification. This would be in line with the approach for sensitivity analyses outlined in technical report \cite{emachmpich4362212017_ich_nodate}. We leave a detailed description of the assumptions and the feasibility of the proposed weighting procedures for future research. 

The proposed weighting procedure also lends a purpose to other parts of the between-within model apart from $\beta^W$. \cite{carlinRegressionModelsTwin2005} offer an interpretation (in a somewhat different setup than ours) of $\beta^B$ assigning meaning to the scenarios $\beta^B \approx 0$ and $\beta^B \approx \beta^W$. It is generally appreciated that the parameter $\beta^B$ is less easily assigned a causal interpretation as it will be confounded also by constant confounders (more precisely, $\beta^B$ pertains to the exposure concordant pairs, but we have only imposed the symmetric confounding assumption ($\widetilde{\text{C1}}$) for the discordant families). This, however is exactly the point of the proposed weighting scheme -- the between family effects are informative about the confounding distribution and may, ideally, be used to estimate it.

Examining the fourth last equality of equation (\ref{eq:12}) we also see that if one could approximate the distribution of $\alpha$ in the target population one could also estimate $\target{3}$ relying on NDI but not SA.  However, this would require estimating the conditional effects given $\alpha$, but models including such interactions are generally not identifiable with sibling pair data barring the observation of variables from $U$.

We have throughout focused on the scenario where there is no effect of a cosibling's exposure on the sibling's exposure. If such an effect was thought to exist, it would be interesting to consider strategies from the literature on estimands \citep[technical report][]{emachmpich4362212017_ich_nodate} for dealing with intercurrent events since, in this scenario, the cosibling's exposure could be argued to constitute such an event, when interference is present. Various modifications to the targets proposed in the present paper could be imagined, for example to estimate the family-level intervention effect $\target{3}$ in a principal stratum defined by the cosibling's exposure \citep[e.g.][]{lipkovichUsingPrincipalStratification2022}. Also note that in this scenario, the target $\target{1}$, in which the cosibling's exposure is left as observed, corresponds to a ``treatment policy'' estimand.

\citeauthor{petersen2020causal} conclude that the causal interpretation of the sibling comparison may be more easily interpretable on the linear scale. We have not made any investigations that directly support this as we have focused on the linear scale. However, casting the NDI and SA assumptions as assumptions of no effect modification highlights the importance of choice of scale, and generally no effect modification on a specific scale will imply its existence on any other scale. Another salient feature of the linear scale is the exact relationship between the conditional estimator and the within effect from the BW model \citep{seaman_review_2014}. On other scales, the two will often be numerically close \citep{neuhaus_between-_1998, neuhaus_separating_2006}, but also see e.g. \cite{brumback_use_2017}, which is why the BW model is sometimes referred to as the ``poor man's'' conditional likelihood. Returning to the connection between the sibling comparison and the cross-over design, \cite{freemanPerformanceTwostageAnalysis1989} describe the importance of additivity of the treatment effect to the cross-over design in order to achieve the variance reduction desired from the design, and \cite{sennGraphicalRepresentationClinical1990} describes the importance of additivity to generalise the trial to a larger population.

Finally, it should be stressed that all considerations in the preceeding manuscript apply when the sibling comparison is playing its absolute home-turf: No non-constant confounding. The impact of non-constant confounding is especially pronounced in models that attempt to infer direction of causation, such as the DoC and ICE Falcon models, as shown, respectively, in \cite{rasmussenMajorLimitationDirection2019} and \cite{sjolanderCautionaryNoteRecently2024}, but these go further than the sibling comparison design considered here, where the directions of effects are presupposed. In practise this assumption will be difficult to justify, but the situation may be salvaged by including further covariates into the model for example to adjust for explanatory variables that are not constant on the family level.

\renewcommand\bibname{References}

\appendix

\chapter{Appendix: Matched comparisons}
\label{sec:append-match-comp}

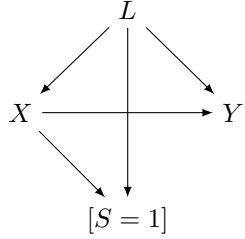
\begin{figure}[htb]
  \centering
  \begin{tikzpicture}
    \node (X) {$X$};
    \node[right = of X] (invis) {};
    \node[right = of invis] (Y) {$Y$};
    \node[above = of invis] (L) {$L$};
    \node[below = of invis] (S) {$[S = 1]$};
    \draw[->] (X) -> (Y);
    \draw[->] (L) -- (X);
    \draw[->] (L) -- (Y);
    \draw[->] (L) -- (S);
    \draw[->] (X) -- (S);
  \end{tikzpicture}
  \caption{Classic matching: DAG representing the assumed relationship between the
    outcomes $Y$, exposures $X$ and confounders $L$. $S$ is an indicator for being sampled (being part of the study) and we have conditioned on being selected ($S = 1$).}
  \label{fig:class-match}
\end{figure}

By classic matching we refer to a scenario where we have the entire group of exposed and wish to compare them to a group of unexposed that is selected so that the two groups match on a set of variables $L$. For example we might for each exposed individual find an unexposed individual so that the two have the same $L$ (one-to-one matching). The classic matching is represented by the DAG in Figure \ref{fig:class-match}. Here $S$ is an indicator for being sampled in the study and we condition on $S=1$ corresponding to being sampled. Both the exposure $X$ and the matching variables $L$ will influence one's probability of being sampled.

This way of matching has two consequences: All exposed will be included in the sample (i.e. $\setdef{A=1} \subseteq \setdef{S = 1}$) and the distribution of $L$ will be the same among exposed and unexposed \emph{in the sample}. Estimation proceeds under the following causal assumptions,
\begin{description}
\item[Consistency] $Y$ equals $Y(x)$ on the event $\setdef{X = x}$
\item[Conditional exchangeability] $Y(x) \indep X \mid L, S = 1$
\end{description}

In our analysis, we compare the mean outcome among exposed and unexposed in our sample, which is an estimate of the following quantity,
\begin{equation}
  \label{eq:34}
  \begin{aligned}
    &\cexpect{Y}{X=1, S=1} - \cexpect{Y}{X=0, S=1} \\
    &=
      \int \cexpect{Y(1)}{X=1, S=1, L = l} f_{L \mid X, S}(l \mid 1, 1)\: dl
      -
      \int \cexpect{Y(0)}{X=0, S=1, L = l} f_{L \mid X, S}(l \mid 0, 1)\: dl
    \\
    &=
      \int \cexpect{Y(1)}{X=1, S=1, L = l} f_{L \mid X, S}(l \mid a, 1)\: dl
      -
      \int \cexpect{Y(0)}{X=0, S=1, L = l} f_{L \mid X, S}(l \mid a, 1)\: dl
    \\
    &=
      \int \cexpect{Y(1)}{L = l} f_{L \mid X, S}(l \mid a, 1)\: dl
      -
      \int \cexpect{Y(0)}{L = l} f_{L \mid X, S}(l \mid a, 1)\: dl
    \\
    &=
      \int
      \left\{
        \cexpect{Y(1) - Y(0)}{L = l}
      \right\}
      f_{L \mid X, S}(l \mid a, 1)\: dl
    \\
    &=
      \int
      \left\{
        \cexpect{Y(1) - Y(0)}{L = l, X= a, S=1}
      \right\}
      f_{L \mid X, S}(l \mid a, 1)\: dl
    \\
    &=
      \cexpect{Y(1) - Y(0)}{X= a, S=1}
    \\
    &=
      \begin{cases}
        \text{ATT} & a = 1 \\
        \cexpect{Y(1) - Y(0)}{X= 0, S=1} & a = 0
      \end{cases}
  \end{aligned}
\end{equation}

where we for the case $a=1$ have used that $\setdef{A=1} \subseteq \setdef{S = 1}$. In particular, we have supplied an argument for the fact, that matching supplies an estimate of the average treatment effect among the treated (ATT). Also note the analoguous situation where we start with the group of unexposed, which would then yield the average treatment effect among the unexposed. As argued in the main text for sibling comparison, if $L$ is not a modifier of the $X$-$Y$ relation, we obtain the ATE, $\expect{Y(1) - Y(0)}$.

\chapter{Appendix: Details of simulations}
\label{cha:append-deta-simul}

Below, we provide details about the simulations presented in Section \ref{sec:simulation-study}.

\section{Data-generating mechanisms}
\label{sec:data-gener-mech}

First, simulate the confounder $U \sim N(0, \sigma_U^2)$ and then simulate,
\begin{equation}
  \label{eq:16}
  \alpha = b_{\alpha} \cdot \exp(b_{\Lambda} \cdot U) + \tilde{\alpha},
\end{equation}
where $\tilde{\alpha} \sim N(0, \tau^2)$. We then simulate the overall family exposure level $\bar{X}_.$ from a multinomial distribution so that,
\begin{equation}
  \label{eq:17}
  \cproba{\bar{X}_. = x}{\alpha} =
  \begin{cases}
    (1 - \Lambda(b_X \cdot \alpha))/2 & \text{ if } x = 0 \\
    \Lambda(b_X \cdot \alpha) & \text{ if } x = 1/2 \\
    (1 - \Lambda(b_X \cdot \alpha))/2 & \text{ if } x = 1
  \end{cases}
\end{equation}
where $\Lambda: x \mapsto [1 + \exp(-x)]^{-1}$ is the logistic function.

The above simulations are performed for $N$ families. We then simulate the sibling-level exposure by setting $X_1 = X_2 = \bar{X}_.$ whenever $\bar{X}_. = 0$ or $\bar{X}_. = 1$ (i.e. in case of exposure concordance). When $\bar{X}_. = 1/2$ (exposure discordance), simulate $X_1 \sim \text{Bernoulli}(\pi_X)$ and set $X_2 = 1 - X_1$. Indexing the families by $i=1,\ldots,N$ and siblings by $j=1,2$ as in the main text, we simulate the outcomes as,
\begin{equation}
  \label{eq:19}
  Y_{ij} = \mu + \alpha_i + \beta^W \cdot X_{i j} + \beta^D \cdot X_{i j} \cdot \alpha_i + \beta^C \cdot X_{i (3 - j)} + \epsilon ,
\end{equation}
with $\epsilon \sim N(0, \sigma^2)$.

\section{Analysis methods}
\label{sec:analysis-methods}

We consider two methods for estimating the target $\target{3} = \expect{Y_1(1,1) - Y_1(0,0)}$.
\begin{enumerate}
\item Fit the between-within model to the selected data (i.e. with $S = 1$)
  \begin{equation}
    \label{eq:47}
    Y_{ij} = \mu + \tilde{\alpha}_i + \beta^B \bar{X}_{i.} + \left(X_{ij} - \bar{X}_{i.}\right) \beta^W + \epsilon_{ij} 
  \end{equation}
  and report $\widehat{\beta^W}$.
\item Weighted estimation
  \begin{enumerate}
  \item Obtain weights in one of three ways.
    \begin{enumerate}
    \item Calculate the true weights from (\ref{eq:17}).
    \item Estimate the weights from two logistic regressions
      \begin{equation}
        \label{eq:50}
        \begin{aligned}
          &M_1: \logit \cproba{X_1 = 1, X_2 = 1}{\alpha} = d_{0,1} + d_{1,1} \alpha \\
          &M_0: \logit \cproba{X_1 = 0, X_2 = 0}{\alpha} = d_{0,0} + d_{1,0} \alpha 
        \end{aligned}
      \end{equation}
      Note that the weight models are misspecified but contain the true weight model. Calculate the weights (note, only varying on family level) as
      \begin{equation}
        \label{eq:49}
        w_{ij} (1) = \indic{\bar{X}_{i.} = 1} \left[\hat{\pi}_1 \right]^{-1}, \quad
        w_{ij} (0) = \indic{\bar{X}_{i.} = 0} \left[\hat{\pi}_0 \right]^{-1}
      \end{equation}
      where $\pi$ is the predicted probability with subscript corresponding to the models from the preceding step. Note that the weights depend on $\alpha$ and are evaluated at the true $\alpha$.
    \item Same as the previous method but using an estimated $\hat{\alpha}$ rather than the true $\alpha$. The estimates are obtained by fitting the between-within model in equation (\ref{eq:47}) and setting,
      \begin{equation}
        \label{eq:36}
        \hat{\alpha}_i = \widehat{\tilde{\alpha}}_i + \hat{\beta}^B \bar{X}_{i.} ,
      \end{equation}
      where $\widehat{\tilde{\alpha}}$ are BLUPs for the family-level effects.
    \end{enumerate}
  \item Estimate the weighted means,
    \begin{equation}
      \label{eq:48}
      \mu(1) = \expect{w_{ij}(1) Y_{ij}}, \quad
      \mu(0) = \expect{w_{ij}(0) Y_{ij}}
    \end{equation}
  \item Report $\widehat{\mu(1)} - \widehat{\mu(0)}$.
  \end{enumerate}
\end{enumerate}

\section{Calculations}
\label{sec:calculations}

Under the simulation model,
\begin{equation}
  \label{eq:20}
  \begin{aligned}
    \cexpect{Y_1 (x_1, x_2)}{\alpha}
    &=
      \cexpect{Y_1}{\alpha, X_1 = x_1, X_2 = x_2}
    \\
    &=
      \mu + \alpha_i + \beta^W \cdot x_1 + \beta^D \cdot x_1 \cdot \alpha + \beta^C \cdot x_2
  \end{aligned}
\end{equation}
and thus,
\begin{equation}
  \label{eq:23}
  \begin{aligned}
    \cexpect{Y_1 (1, 0)}{\alpha} - \cexpect{Y_1 (0, 1)}{\alpha}
    &=
      \beta^W + \beta^D \cdot \alpha - \beta^C 
  \end{aligned}
\end{equation}
and we see that the (SA) assumption corresponds to $\beta^D = 0$. Further,
\begin{equation}
  \label{eq:24}
  \begin{aligned}
    \cexpect{Y_1 (1, 1)}{\alpha} - \cexpect{Y_1 (0, 0)}{\alpha}
    &=
      \beta^W + \beta^D \cdot \alpha + \beta^C
  \end{aligned}
\end{equation}
Comparing (\ref{eq:23}) with (\ref{eq:24}) we see that the (NDI) assumption corresponds to $\beta^C = 0$. Finally,
\begin{equation}
  \label{eq:32}
  \begin{aligned}
    \target{3}
    &=
      \expect{\cexpect{Y_1 (1, 1)}{\alpha} - \cexpect{Y_1 (0, 0)}{\alpha}} \\
    &=
      \beta^W + \beta^D \cdot  b_{\alpha} \cdot \exp(b_{\Lambda}^2 \sigma_U^2/2) + \beta^C
  \end{aligned}
\end{equation}

\section{Scenarios and parameters}
\label{sec:scenarios-parameters}

The parameters for the simulations are given in Table \ref{tab:sim-params}.

\begin{table}[htb]
  \centering
\begin{tabular}{rrrrr}
  \hline
 & Scenario 1 & Scenario 2 & Scenario 3 & Scenario 4 \\ 
  \hline
$N$ & 1000.00 & 1000.00 & 1000.00 & 1000.00 \\ 
  $b_{\Lambda}$ & 0.50 & 0.50 & 0.50 & 0.50 \\ 
  $\sigma_U$ & 2.00 & 2.00 & 2.00 & 2.00 \\ 
  $b_{\alpha}$ & 0.40 & 0.40 & 0.40 & 0.40 \\ 
  $b_{X}$ & 0.20 & 0.20 & 0.20 & 0.20 \\ 
  $\pi_X$ & 0.50 & 0.50 & 0.50 & 0.50 \\ 
  $\tau$ & 2.00 & 2.00 & 2.00 & 2.00 \\ 
  $\mu$ & 10.00 & 10.00 & 10.00 & 10.00 \\ 
  $\beta^B$ & 2.00 & 2.00 & 2.00 & 2.00 \\ 
  $\beta^W$ & 5.00 & 5.00 & 5.00 & 5.00 \\ 
  $\beta^D$ & -1.00 & 0.00 & -1.00 & 0.00 \\ 
  $\beta^C$ & 1.50 & 1.50 & 0.00 & 0.00 \\ 
  $\sigma$ & 1.00 & 1.00 & 1.00 & 1.00 \\ 
   \hline
\end{tabular}

  \caption{Simulation parameters}
  \label{tab:sim-params}
\end{table}

\chapter{Appendix: A second weighting procedure}
\label{sec:append-weight-proc}

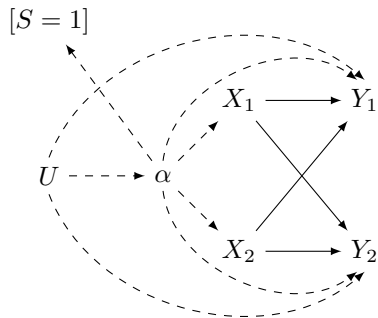
\begin{figure}
  \centering
  \begin{tikzpicture}
    \node[] (U) {$U$};
    \node[right = of U] (a) {$\alpha$};
    \node[above = 1.5cm of U] (S) {$\left[S = 1\right]$};
    \node[right = of a, above of = a] (X1) {$X_1$};
    \node[right = of a, below of = a] (X2) {$X_2$};
    \node[right = of X1] (Y1) {$Y_1$};
    \node[right = of X2] (Y2) {$Y_2$};
    \draw[->] (X1) -> (Y1);
    \draw[->] (X2) -> (Y2);
    \draw[->] (X1) -> (Y2);
    \draw[->] (X2) -> (Y1);
    \draw[->, dashed] (U) -- (a);
    \draw[->, dashed] (a) -- (X1);
    \draw[->, dashed] (a) -- (X2);
    \draw[dashed] (a.90) edge[bend left = 60, ->] (Y1.110);
    \draw[dashed] (a.270) edge[bend right = 60, ->] (Y2.250);
    \draw[dashed] (a) edge[->] (S);
    \draw[dashed] (U.90) edge[bend left = 55, ->] (Y1.90);
    \draw[dashed] (U.270) edge[bend right = 55, ->] (Y2.270);
  \end{tikzpicture}
  \caption{Slight simplification of the DAG in Figure \ref{fig:dag-potential} under which the weighting procedure is analysed. The simplification consists of the assumption that $U$ affects the selection $S$ only through $\alpha$.}
  \label{fig:dag-weight}
\end{figure}

Consider the DAG in Figure \ref{fig:dag-weight}, which is a slight simplification of that in Figure \ref{fig:dag-potential}, where we have removed the arrow from $U$ to $S$, corresponding to the assumption that the confounders affect the selection only through $\alpha$. 

Consider the set of weights defined in equation (\ref{eq:15}) of the main text,
\begin{equation}
  \label{eq:38}
  w(a, x_1, x_2) = \indic{X_1 = x_1, X_2 = x_2, S = 1} \left[f_{(X_1, X_2) \mid \alpha} (x_1, x_2 \mid a) f_{S \mid \alpha} (1 \mid a) \right]^{-1}
\end{equation}
We note from the DAG that $(X_1, X_2)$ is conditionally independent from $S$ given $\alpha$ and $U$, and thus, \begin{equation}
  \label{eq:40}
  \begin{aligned}
    \expect{w(\alpha, x_1, x_2) Y}
    &=
      \int
      \left[f_{(X_1, X_2) \mid \alpha} (x_1, x_2 \mid a) f_{S \mid \alpha} (1 \mid a) \right]^{-1} \\
    &\qquad\qquad \cexpect{\indic{X_1 = x_1, X_2 = x_2, S = 1} Y}{\alpha = a, U = u}
      f_{(\alpha, U)} (a, u)
      \: (da, du) \\
    &=
      \int
      \frac{
      f_{(X_1, X_2, S) \mid \alpha, U} (x_1, x_2, 1 \mid a, u)
      }{
      f_{(X_1, X_2) \mid \alpha} (x_1, x_2 \mid a) f_{S \mid \alpha} (1 \mid a)
      } \\
    &\qquad\qquad
      \cexpect{Y}{\alpha = a, X_1 = x_1, X_2 = x_2, S = 1, U = u}
      f_{(\alpha, U)} (a, u)
      \: (da, du) \\
    &=
      \int
      1 \cdot
      \cexpect{Y}{\alpha = a, X_1 = x_1, X_2 = x_2, S = 1, U = u}
      f_{(\alpha, U)} (a, u)
      \: (da, du) \\
    &=
      \int
      \cexpect{Y(x_1, x_2)}{\alpha = a, X_1 = x_1, X_2 = x_2, S = 1, U = u}
      f_{(\alpha, U)} (a, u)
      \: (da, du) \\
    &=
      \int
      \cexpect{Y(x_1, x_2)}{\alpha = a, U = u}
      f_{(\alpha, U)} (a, u)
      \: (da, du) \\
    &=
      \expect{Y(x_1, x_2)}
  \end{aligned}
\end{equation}

Since $\alpha$ is not observed one would in practise evaluate the weight at the BLUP and use the weight $w(\hat{\alpha}, x_1, x_2)$, see the Discussion in the main paper. Equation (\ref{eq:40}) shows that this weighting procedure allows estimation of any of the targets and does not rely on NDI or SA.

\end{document}